# SoNDe

## Solid-State Neutron Detector

### INFRADEV-1-2014/H2020

### Grant Agreement Number: 654124

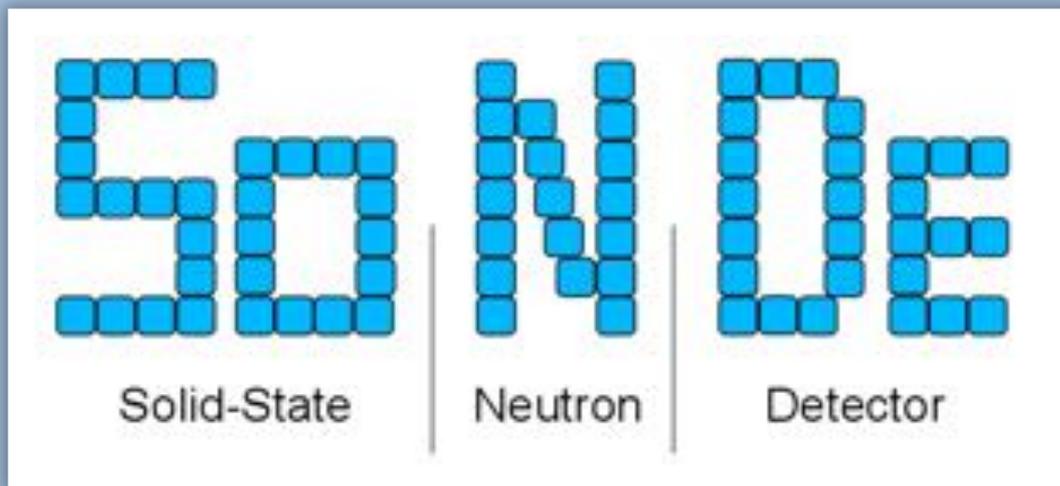

### Deliverable Report: D1.1 Demonstrator

**This project is funded by the Horizon 2020 Framework Programme of the European Union. Project number 654124.**

# Project and Deliverable Information Sheet

| SoNDe Project | Project Ref. No. 654124 | |
|---|---|---|
| | Project Title: Solid-State Neutron Detector SoNDe | |
| | Project Website: http://www.fz-juelich.de/ics/ics-1/DE/Leistungen/ESS/SoNDe-Projekt/ | |
| | Deliverable ID: D1.1 | |
| | Deliverable Nature: 1x1 Hardware Demonstrator | |
| | Deliverable Level: PU | Contractual Date of Delivery: 31.07.2015 |
| | | Actual Date of Delivery: 05.08.2015 |
| | EC Project Officer: Bernhard Fabianek | |

# Document Control Sheet

| Document | Title: Solid-State Neutron Detector SoNDe | |
|---|---|---|
| | ID: Website-Deliverable-D1.1 | |
| | Version: 1.0 | |
| | Available at: | |
| | Software Tool: MS Word 2011 | |
| | Files: 1x1-Demo-Deliverable-D1.1.docx | |
| Authorship | Written by | Sebastian Jaksch, FZJ |
| | Contributors | Ralf Engels, Günter Kemmerling (FZJ) |
| | Reviewed by | Sebastian Jaksch, FZJ |
| | Approved | Sebastian Jaksch, FZJ |

# List of Abbreviations

| FZJ | Forschungszentrum Jülich, Jülich Research Centre |
| JCNS | Jülich Centre for Neutron Science |
| LLB | Laboratoire Léon-Brillouin |
| ESS | European Spallation Source |
| IDEAS | Integrated Detector Electronics AS |
| MaPMT | Multi-anode Photomultiplier Tube |
| ROSMAP | IDEAS code name for counting electronics |

## List of Figures

This project is funded by the Horizon 2020 Framework Programme of the European Union. Project number 654124.

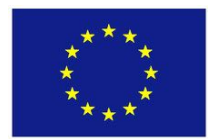

# Table of Contents

**Executive Summary**

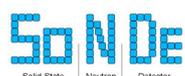

**This project is funded by the Horizon 2020 Framework Programme of the European Union. Project number 654124.**

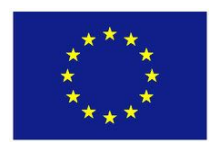

This 1x1 demonstrator shows a single module of the envisaged detector technology, which will lead to the construction of a large scale detector during the course of the SoNDe project.

This is both to show the feasibility and the capabilities of the proposed technology. The demonstrator will also allow for intensive testing to learn about any issues connected with upscaling or technical feasibility. To perform these tests at an early stage of the project allows for finding and correcting issues which, if found later during the course of the SoNDe project might lead to major delays and/or technical or budgetary problems. Thus building the 1x1 demonstrator is part of the risk avoidance strategy within the SoNDe project.

The demonstrator consists of a pixelated scintillator/carrier glass sandwich, a H8500 Hamamatsu MaPMT (Multi-anode Photomultiplier Tube) and a ROSMAP counting chipsystem to read out the MaPMT signals. All of these are mounted in series and the neutron will be detected as it approaches from the glass sandwich side.

## Introduction

One of the main goals in the first workpackage WP1 (prestudies) is to determine technical feasibility and capabilities of the proposed technology. This is done in order to avoid later issues with the technology and to find the optimal solution by extensive testing of a 1x1 demonstrator. Both goals are served by creating said demonstrator.

The main issues identified at this point of the project are the reliable cutting or milling of the pixelated scintillator glass placed onto the carrier glass, connection between readout electronics and MaPMT and reliability of the software framework for further testing of the demonstrator.

All these three issues have been addressed and solved during the construction of the 1x1 demonstrator. The cutting of the scintillator glass can be performed reliably by a wafer saw if used at subsequential low cutting depth. For the connection between the readout electronics and the MaPMT a programmable FPGA chip is the most flexible solution as it allows for later adaption of the electronics to specific needs that may arise during the development of the project. Finally, the software to control the measurements was deemed inappropriate for final use in a large-scale detector, however a software development was anticipated in the upscaling part of the project (WP 3) and will be part of the integration into a working environment (deliverable D3.3). The experiences made with the first software, which will be continuously updated, will then feed into that result.

## Construction of the 1x1 demonstrator

As previously described the 1x1 demonstrator consists of a pixelated scintillator/carrier glass sandwich, a H8500 Hamamatsu MaPMT (Multi-anode Photomultiplier Tube) and a ROSMAP counting chipsystem to read out the MaPMT signals. All of these are mounted onto a frame for later testing and ease of handling. The construction of each of these components of the demonstrator, which in itself is a first prototype for the final modules to be used in the large scale detector later, will be explained in the following, each in a dedicated section.

This project is funded by the Horizon 2020 Framework Programme of the European Union. Project number 654124.

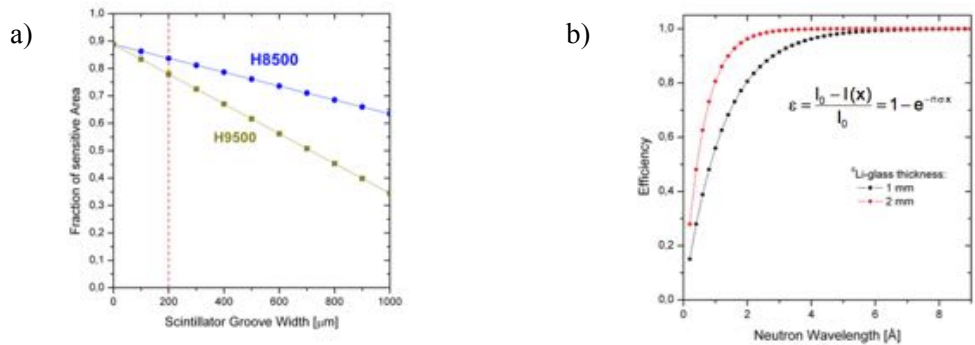

**Fig. 1. a)** Calculated sensitive area for H8500 and H9500 MaPMTs as a function of cutting width and **b)** calculated efficiency as a function of glass thickness and neutron wavelength. From a) it is immediately visible that as little dead space, i.e. as much as possible sensitive area can be achieved by using a small groove width, leading to the requirement of a slim saw. In b) it becomes clear that at the usual employed wavelengths in neutron scattering instruments of 4 Å and above both the 1 and 2 mm scintillator have virtually the same efficiency.

## Pixelated Scintillator/Carrier glass sandwich

In order to be able to have a pixelated scintillator with as little dead space, i.e. non scintillating area, the most feasible way to attach the scintillator glass as a full pane to a carrier glass and then cut grooves into it (see Fig. 1). The other option of single scintillation pixels which are afterwards fixed onto the MaPMT both requires a lot more assembly work and the minimally achievable distance between pixels is larger. With the construction of a glued sandwich, where the scintillator glass is afterwards grooved by a saw, mechanical stability, minimal dead space and ease of a assembly are combined in an ideal way.

First attempts at grooving the glass with a saw resulted in chipped edges, as the mechanical stress to the scintillator glass was to high (see Fig. 2 a). This issue was resolved by cutting subsequentially for several times with a maximal cutting depth in the order of 50 µm. Thus the mechanical stress could be reduced far enough that chipping of the edges of the scintillator did not occur while maintaining a maximal with of the cuts of 250 µm (See Fig. 2 b).

## H8500 Hamamatsu MaPMT

The H8500 MaPMT was purchased from Hamamatsu (See Fig. 3). No further modification were made. The glass sandwich can be positioned in front of the MaPMT either with glue or optical gel.

## ROSMAP counting electronics

The ROSMAP counting electronics (see Fig. 4) and software were purchased from IDEAS. Up to this point no further adaptions were made. Connections between the H8500 and the electronics are made by pinhead connectors.

This project is funded by the Horizon 2020 Framework Programme of the European Union. Project number 654124.

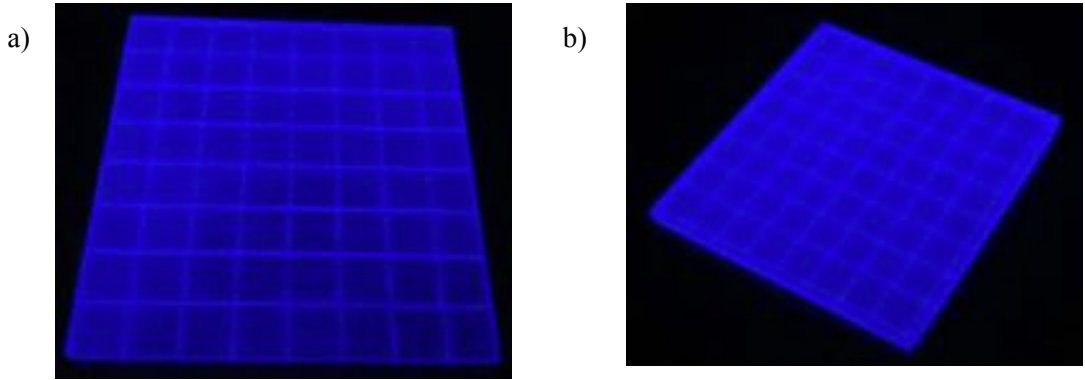

**Fig. 2. False color photographs of test samples where the scintillator glass was cut with a) a common glass saw and b) a wafer saw. False colors are used for better visibility of the edges. While cuts with the glass saw resulted in chipped edges (e.g. see second row from the bottom in a), cutting with a wafer saw resulted in smooth edges of the scintillator glass while maintaining thin cutting lines.**

## Conclusion

A working 1x1 demonstrator was presented (see Fig. 4). This demonstrator will later be used for further testing under working conditions in neutron scattering environments. Moreover, the choice of types of different material can be facilitated by testing any possible alternatives in this module prototype. This prototype shows the feasibility of the proposed technique while the capabilities will be shown during further testing of this demonstrator.

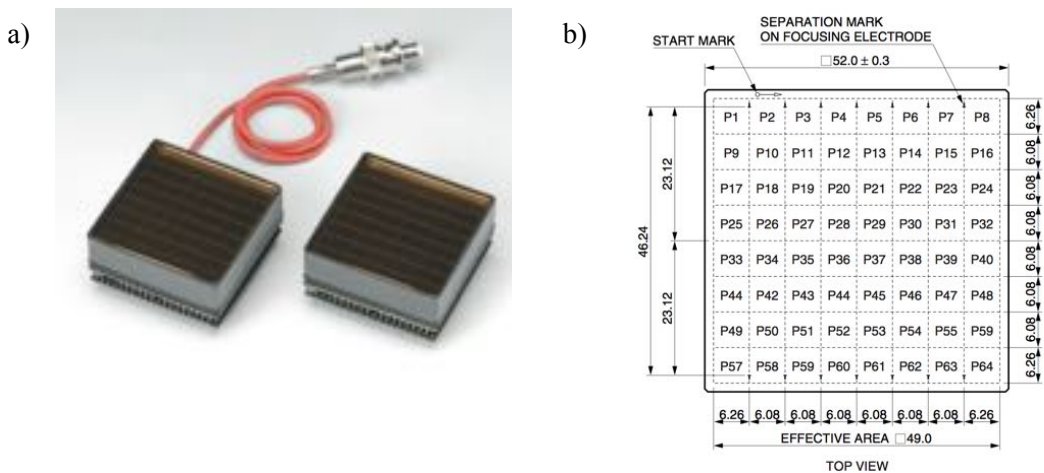

**Fig. 3. a) photograph of a H8500 MaPMT and b) technical sketch of the same MaPMT. (Images from Hamamatsu Technical Manual, FLAT PANEL TYPE MULTIANODE PMT ASSEMBLY H8500 SERIES / H10966 SERIES, Hamamatsu Photonics Inc., Shimokanzo, Japan.)**

a)                                                    b)

This project is funded by the Horizon 2020 Framework Programme of the European Union. Project number 654124.

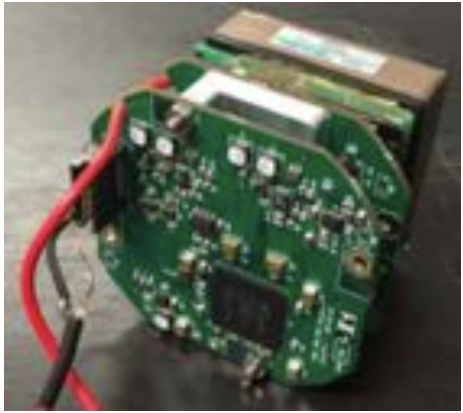 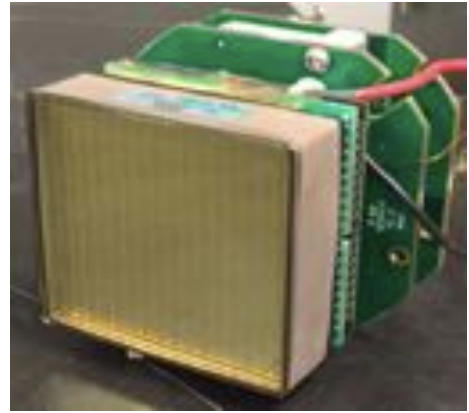

**Fig. 4. Photograph of a) back and b) front of the assembled 1x1 demonstrator. Here the glass sandwich is removed for protective reasons and better visibility. In a) the ROSMAP is visible as a stack from behind. The circuit boards are nearly completely covered by the MaPMT when seen from the front. Further compactification of the board will make this possible and thus achieve the goal of hiding the complete electronics behind the MaPMTs and not creating additional dead space.**

**This project is funded by the Horizon 2020 Framework Programme of the European Union. Project number 654124.**

# SoNDe

## Solid-State Neutron Detector

### INFRADEV-1-2014/H2020

### Grant Agreement Number: 654124

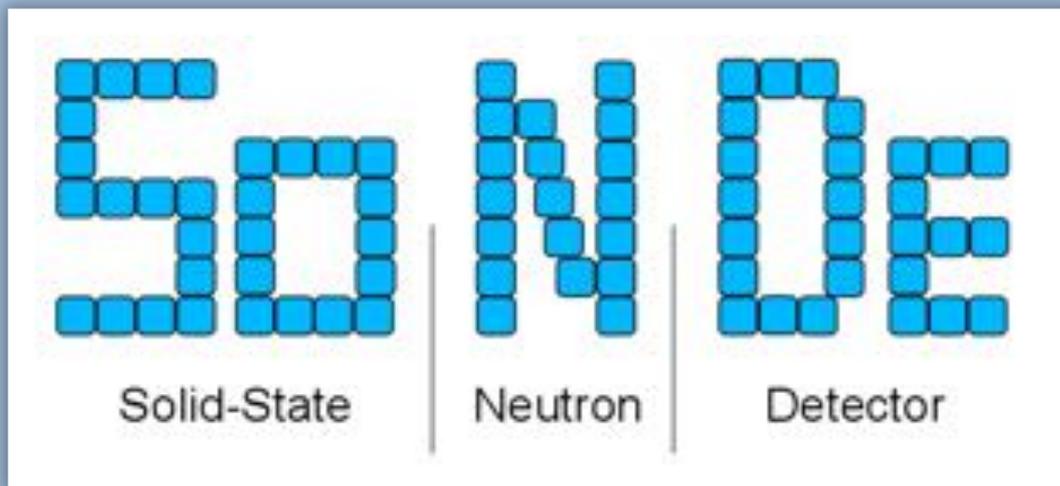

Deliverable Report: D1.2 Materials Report

**This project is funded by the Horizon 2020 Framework Programme of the European Union. Project number 654124.**

## Project and Deliverable Information Sheet

| SoNDe Project | |
|---|---|
| | Project Ref. No. 654124 |
| | Project Title: Solid-State Neutron Detector SoNDe |
| | Project Website: http://www.fz-juelich.de/ics/ics-1/DE/Leistungen/ESS/SoNDe-Projekt/ |
| | Deliverable ID: D1.2 |
| | Deliverable Nature: Materials Report |
| | Deliverable Level: PU |

| | Contractual Date of Delivery: 31.10.2015 |
|---|---|
| | Actual Date of Delivery: 24.11.2015 |

| | EC Project Officer: Bernhard Fabianek |
|---|---|

## Document Control Sheet

| Document | |
|---|---|
| | Title: Solid-State Neutron Detector SoNDe |
| | ID: MaterialsReport-D1.2 |
| | Version: 1.0 |
| | Available at: |
| | Software Tool: MS Word 2011 |
| | Files: MaterialsReport-Deliverable-D1.2.docx |

| Authorship | Written by | Günter Kemmerling, Sebastian Jaksch, FZJ |
|---|---|---|
| | Contributors | |
| | Reviewed by | Sebastian Jaksch, FZJ |
| | Approved | Sebastian Jaksch, FZJ |

## List of Abbreviations

This project is funded by the Horizon 2020 Framework Programme of the European Union. Project number 654124.

| | |
|---|---|
| FZJ | Forschungszentrum Jülich, Jülich Research Centre |
| H8500 | Type of MaPMT used in the SoNDe project |
| JCNS | Jülich Centre for Neutron Science |
| LLB | Laboratoire Léon-Brillouin |
| ESS | European Spallation Source |
| IDEAS | Integrated Detector Electronics AS |
| MaPMT | Multi-anode Photomultiplier Tube |
| ROSMAP | IDEAS code name for counting electronics |
| TOF | time-of-flight |

## List of Figures

## Table of Contents

This project is funded by the Horizon 2020 Framework Programme of the European Union. Project number 654124.

## Executive Summary

This report details the process for finding feasible materials for the use within the SoNDe project, focusing on the scintillation materials, which convert the neutron into detectable light, and the used photomultipliers for the detection of said light.

Several alternatives are considered and the choice for the current prototype is rationalized. The current prototype is using $^6$Li-glass as a scintillator and a Hamamatsu H8500 as a photomultiplier.

## Introduction

Within the SoNDe-project a neutron detector based on a solid-state scintillator is to be developed. Such detectors have already been in use for a long time and offer some attractive properties. The relatively high density of the scintillation material allows for thin neutron converters with high detection efficiency and a well-defined interaction point of the neutron. As a result of the low thickness, scintillation detectors also have almost no parallaxes effects and usually offer a high time resolution for the neutron capture.

The special characteristic of the neutron detector in SoNDe is the ability to cope with extremely high count rates of about 20 MHz at 10% dead time on a 1 m$^2$ detector area, associated with a position resolution of 6 x 6 mm$^2$ or 3 x 3 mm$^2$. In order to achieve these goals, a proper selection of the scintillator material and the light sensor is necessary, which is discussed in this report.

## Solid-State Scintillators for thermal neutron detection in SoNDe

Today, there are essentially two types of solid state scintillators used for the detection of thermal neutrons, Li-glass and LiF/ZnS scintillators. In both types of scintillators the Lithium content is enriched with the isotope $^6$Li, which offers a high cross section for neutron capture via the reaction

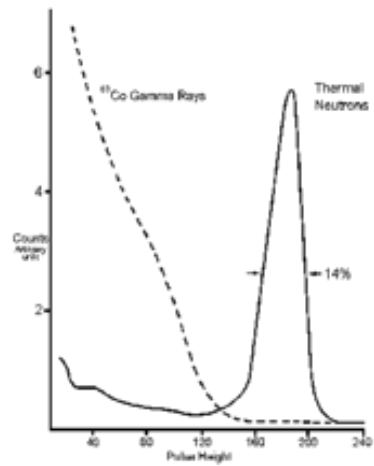

Fig. 1. Pulse height spectrum of Li-glass scintillator showing the pulse heights from neutron capture relative to that of $^{60}$Co-gamma rays (brochure from scintacor, formerly AST).

$$n + {}^6Li \rightarrow \alpha + {}^3H + 4.78\ MeV$$

The secondary particles of this reaction lose their energy within the surrounding scintillator material in close distances to the interaction point. This gives rise to the emission of light, which is to be detected with a light sensor.

Li-glass scintillators are described in [1]-[4] and are available in different types. For thermal neutron detection GS20 type glass scintillators are favorable, which contain 6.6 weight-percent of the $^6$Li isotope. This high $^6$Li content offers already with 1 mm thickness a detection efficiency of about 75% for thermal neutrons. GS20 glass

This project is funded by the Horizon 2020 Framework Programme of the European Union. Project number 654124.

scintillators are doped with Cerium as an activator, which mainly determines the properties of light emission. There are up to 6600 photons emitted per neutron capture ([5]) with a peak maximum at 390 nm wavelength. The decay time of the light emission is very fast, about 60 ns. As the material is transparent, the full amount of photons can be measured outside with an adequate light sensor. The gamma sensitivity of GS20 Li-glass is strongly dependent on the gamma energy; for gamma energies of less than 1 MeV the detection of gammas can be suppressed by pulse height discrimination methods, but for higher energies it is an issue, as these gammas cannot be distinguished from neutrons.

Ceramic scintillators based on a mixture of LiF and ZnS are also often used for neutron detection ([6]-[8]). These LiF/ZnS-scintillators are usually doped by Ag as an activator and have a very high light yield of up to 180000 photons per neutron capture with a peak maximum at 450 nm. Unfortunately both major components are opaque. Thus, due to the self-absorption of the emitted light the thickness is limited to 450-500 μm, which impacts detection efficiency. Another drawback of such scintillators is the rather high decay time of the light emission, which is a few μs and which limits the count rate capability. The gamma sensitivity can be reduced to $10^{-5}$ by pulse shape discrimination methods ([9]).

The high count-rate capability to be achieved in SoNDe requires a short signal processing time. Pile-up effects, which occur due to overlapping signals have to be avoided as this behavior hinders a discrimination between neutrons and gammas or makes it even impossible. As the length of the output pulses of the light sensor is closely related to the length of the light emission of the scintillator, the decay time of the scintillator has a high impact on the achievable high count-rates. In this respect the GS20 Li-glass scintillator, with a decay time faster by about a factor of 20, is clearly advantageous over the LiF/ZnS scintillator. This is also true related to the detection efficiency. As the LiF/ZnS scintillator is opaque, its thickness is limited due to light detection constraints. Since the number of neutron capture is dependent on the thickness, the detection efficiency is thereby also limited. As long as the environment of the detector doesn't show gammas with energies higher 1 MeV, the GS20 Li-glass scintillator also provides a simple and fast discrimination between neutrons and gammas by pulse height selection. Thus, the properties of the GS20 Li-glass scintillator seem to be superior over the LiF/ZnS scintillator to achieve the goals given in the SoNDe-project.

## Light Sensor Device

The light sensor for the SoNDe detector should provide a fast and efficient detection the scintillator light and allow for the reconstruction of position information with a resolution of 6 x 6 mm$^2$ or 3 x 3 mm$^2$.

There are different methods to extract the position information from the light created by the neutron capture of the scintillator. A common method is based on the Anger principle [10], which is shown in Fig. 2.

This project is funded by the Horizon 2020 Framework Programme of the European Union. Project number 654124.

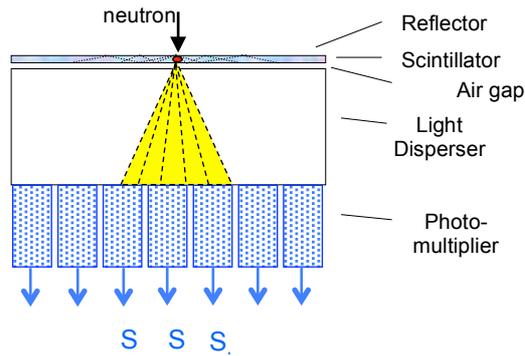

The light created by the neutron capture within the scintillator is spread over several light sensor devices. By analyzing the light sensor signals, the position of the neutron capture can be calculated with a precision, which is much lower than the size of the light sensor devices. Unfortunately, the Anger method requires the determination of the light sensors that are hit by the emitted light and a digitization of their analog signals for the position reconstruction. This is typically a rather time consuming process, which limits the count rate capability of such detectors. Furthermore, the Anger method always occupies the area of several light sensors by one neutron capture. A much faster approach is the usage of a pixelated light sensor structure, where only one pixel is involved per neutron event. This method is realized within the SoNDe project. Together with the selection of a scintillator, which allows for a neutron/gamma discrimination by setting a comparator threshold, the count rate capability of the detector can be improved drastically.

In order to realize a pixelated sensor structure, currently only multianode photomultiplier (MaPMT) and arrays of silicon photomultiplier (SiPM) are the only feasible options. Single photomultiplier tubes or single SiPM modules usually need a larger guard space around the sensitive area, which decreases the overall neutron sensitive area and thereby also the overall detection efficiency.

MaPMTs are based on the common light detection method of vacuum tubes with a photosensitive cathode and several dynode stages, at which electron acceleration by electrical fields is used to yield a multiplication of the initial charge. SiPMs are based on a microstructure of photosensitive semiconductor diodes, which are operated in the breakthrough mode, such that an absorbed photon results in a charge avalanche at the anode of the diode (avalanche photodiode).

Both, MaPMTs and SiPMs, are available with a sufficient sensitivity to 390 nm light and both types provide also pixelated sensor structures of 6 mm x 6 mm or 3 mm x 3 mm. With respect to magnetic fields, SiPMs are much less sensitive but their gain shows a larger dependence on temperature changes. Anyhow, there are publications ([12]) showing that SiPMs undergo radiation damage effects with thermal neutrons at integrated doses, which are considerably lower than instrument life times at neutron sources (at the ESS for a SANS instrument approximately $5 \times 10^{14}$ neutrons/cm$^2$ are expected over a lifetime of 10 years). Currently, this is a very important disadvantage of SiPMs which rules out their usage for neutron detection.

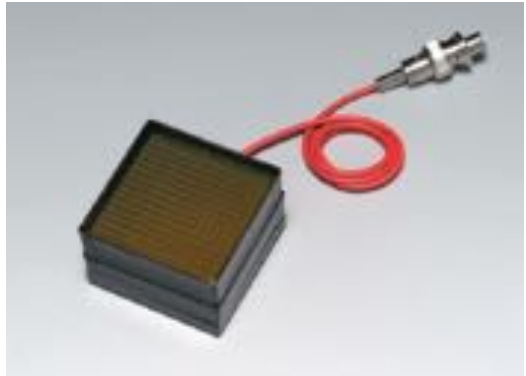

There are two types of MaPMTs produced by Hamamatsu, H8500 and H9500, which are suitable solutions for the SoNDe detector type. Both models have an overall size of 52 mm x 52 mm with only a small guard ring around the sensitive area such that a sensitive area of 89% can be achieved. Due to the small dead space, this MaPMTs offer scalable and modular detector designs with the possibility to build large detector areas by stacking several modules close together.

Also the electrical properties of the H8500 and H9500 fit well with high rate applications. The overall maximum current for a MaPMT anode is 100 μA. Taking into account that the MaPMTs typically are operated at a gain of $10^6$ and that a neutron event will deliver about 400-500 photons per pixel, this should allow for rates of more than 1 MHz per module.

MaPMTs or devices with similar capabilities from other suppliers, such as the PLANACON photodetector [13], have been considered. However in any case there were technical reasons, which rendered the devices unsuitable for being used in the SoNDe detector concept. In the case of the PLANACON photodetector the readout time was to large to accommodate the needed high time resolution.

## Conclusion

In this report we rationalized and detailed the choices made for the scintillator material as well as for the used MaPMT.

In case of the scintillator material we decided to use $^6$Li-glass due to the good optical properties, the easy handling as well as the good neutron conversion characteristics. This allows us to have a material, where the maximum output of photons to the MaPMT is achieved and in parallel to have a good neutron/gamma discrimination.

The MaPMT was chosen due to its high native resolution combined with a high count-rate capability. Moreover, PMTs have shown little to no degradation under high neutron fluxes, leading to an improved lifetime. Also, by its high time-resolution, this type of MaPMT allows for a very sensitive TOF neutron detector.

**This project is funded by the Horizon 2020 Framework Programme of the European Union. Project number 654124.**

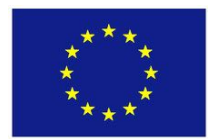

# SoNDe

## Solid-State Neutron Detector

### INFRADEV-1-2014/H2020

### Grant Agreement Number: 654124

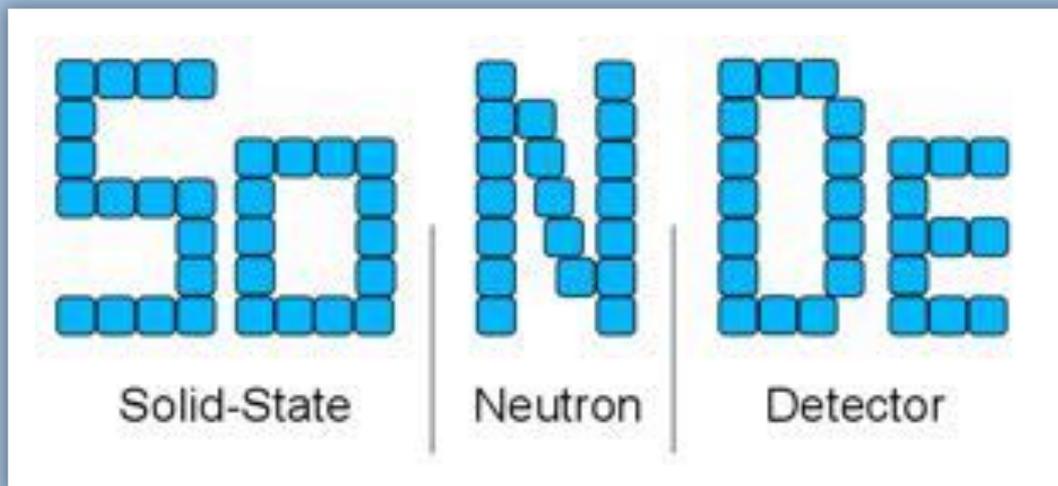

Deliverable Report: D1.3 Electronics Report on Radiation Hardness

This project is funded by the Horizon 2020 Framework Programme of the European Union. Project number 654124.

## Project and Deliverable Information Sheet

| SoNDe Project | | |
|---|---|---|
| | Project Ref. No. 654124 | |
| | Project Title: Solid-State Neutron Detector SoNDe | |
| | Project Website: http://www.fz-juelich.de/ics/ics-1/DE/Leistungen/ESS/SoNDe-Projekt/ | |
| | Deliverable ID: D1.3 | |
| | Deliverable Nature: D1.3 Electronics Report on Radiation Hardness | |
| | Deliverable Level: PU | Contractual Date of Delivery: 01.10.2015 |
| | | Actual Date of Delivery: 14.12.2015 |
| | EC Project Officer: Bernhard Fabianek | |

## Document Control Sheet

| Document | | |
|---|---|---|
| | Title: D1.3 Electronics Report on Radiation Hardness | |
| | ID: D1.3 Electronics Report on Radiation Hardness - Deliverable-D1.3 | |
| | Version: 1.0 | |
| | Available at: http://www.fz-juelich.de/ics/ics-1/EN/Leistungen/ESS/SoNDe-Projekt/Reports/_node.html | |
| | Software Tool: MS Word 2011 | |
| | Files: D1.3 Electronics Report on Radiation Hardness.docx | |
| Authorship | Written by | Codin Gheorghe, IDEAS |
| | Contributors | Philip Påhlsson, IDEAS, Sebastian Jaksch, FZJ |
| | Reviewed by | Sebastian Jaksch, FZJ |
| | Approved | Sebastian Jaksch, FZJ |

## List of Abbreviations

| | |
|---|---|
| FZJ | Forschungszentrum Jülich, Jülich Research Centre |
| JCNS | Jülich Centre for Neutron Science |
| LLB | Laboratoire Léon-Brillouin |
| DUT | Device Under Test |
| ESS | European Spallation Source |
| IDEAS | Integrated Detector Electronics AS |
| MaPMT | Multi-anode Photomultiplier Tube |
| ROSMAP | IDEAS code name for counting electronics |
| SEE | Single Event Effect |
| SEL | Single Event Latch-Up |
| SEU | Single Event Upset |
| SET | Single Event Transient |
| SEGR | Single Event Gate Rupture |
| SEB | Single Event Burn Out |
| TID | Total Ionizing Dose |

# List of Figures

# List of Tables

This project is funded by the Horizon 2020 Framework Programme of the European Union. Project number 654124.

# Contents

This project is funded by the Horizon 2020 Framework Programme of the European Union. Project number 654124.

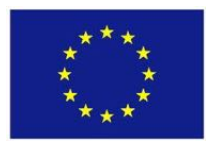

# Executive summary

The ROSMAP module has been tested in at the LLB cold neutron beam in Saclay, France.

The module was subjected to a large set of beam runs. The total accumulated neutron dose is estimated to be more then $1 \times 10^{14}$ neutrons/cm$^2$. No errors in the SoNDe prototype module is observable up to $5 \times 10^{13}$ neutrons/cm$^2$, where communication to the module is lost. The cause of the failure is not known. No SEE or TID effects are visible in the obtained data. The accumulated rate at failure exceeded the rate expected at other sources (e.g. European Spallation Source (ESS)) over a period of approximately 10 years.

The SoNDe prototype was tested in IDEAS lab facilities after the irradiation campaign. Lab tests show that the prototype is functional after irradiation.

# 1 Introduction

## 1.1 Radiation Effects on Electronics

Radiation effects in semiconductor electronics can be categorized as follows

- Single Event Effects (SEE)
- Cumulative effects
- Displacement effects

Single event effects can be considered the localized effect inside a Si-SiO$_2$ junction due to direct ionization or recoiling nuclei from a nuclear interaction by one energetic particle traversing the device. In high-flux environments multiple particles may traverse the device within the time for diffusion causing a superposition effect. Majority of SEU effects are based on creation or separation of charge carriers inside the semiconductor material due to drift, diffusion or funneling. [RD02]

Single event effects can be further categorized into

- Single event latch-up (SEL)
- Single event upset (SEU)
- Single event transient (SET)
- Single event gate rupture (SEGR)
- Single event burn out (SEB)

Single event upsets (SEU) affect memory cells by charge accumulation close to critical volumes, that causes the gate voltage on for example a nMOS transistor to exceed the gate-source threshold voltage, creating a conductive path from drain to source. This can cause a memory element to shift its stored value from a '0' to a '1' or vice versa. This causes memories to change contents in response to impinging radiation. If the radiation has high LET, each ionization track can affect multiple bits causing multiple bit errors (MBU). A common way to prevent SEU errors is to design memories with redundancy either in critical paths or in entire memory elements, creating so called triple modular redundant (TMR) memory elements. The output of TMR latches or flip-flops are routed to voting logic that evaluates all three memory element values and in case of mismatch can flag or correct the erroneous bit value.

SEU is a soft error meaning that it does only require a rewrite of the bit in order to clear.

This project is funded by the Horizon 2020 Framework Programme of the European Union. Project number 654124.

Single event latch-up (SEL) is a permanent and potentially destructive triggering of a parasitic PNPN thyristor that is caused by energetic particle interaction with the ASIC. A PNPN structure results in an uncontrolled increase of component supply current, which might subsequently lead to component destruction (burnout).

Ionizing particle-induced latch-up is a possibly destructive phenomenon for circuits operating in radiation environments. An ionising particle can generate a large number of electron-hole pairs along its penetration track and if sufficient ionising charge is generated, it can cause a latch-up. The charge does not directly latch-up the device but it occurs in a number of steps. Seeing the low charge generation of heavy ionising particles, the area of interaction is of importance. The main part of the charge causing latch-up is generated in a sensitive area of the device in the well-substrate junction. In order for latch-up to occur, the transient current form the ion strike in this area must be sufficient to forward bias the parasitic vertical PNP transistor, see Figure 1. The amplified current from the vertical PNP transistor, if sufficient, will turn on the lateral NPN transistor causing a sustained bias condition for the vertical PNP transistor. The positive feedback draws excessive current that is only limited by the connected load or power supply. SEL is in effect a short circuit of the ASIC power supply and if the ASIC power supply is not removed fast enough can lead to circuit breakdown. In order to prevent latch-up the parasitic transistor loop gain and resistance from the all circuit points to the power supplies shall be minimized.

The parasitic transistors, if switched on can only be switched off by removing the ASIC supply voltage. To prevent SEL ASICs can be designed with guard rings around the NMOS and PMOS transistors in order to collect injected majority and minority carriers close to the parasitic PNP and NPN transistors, thus preventing them from entering a forward biased state. The guard rings are expected to divert the majority of the injected charge and hence mitigate the risk of SEL. The addition of guard rings lower the substrate and well resistance, hence lowering the parasitic transistor loop gain.

This project is funded by the Horizon 2020 Framework Programme of the European Union. Project number 654124.

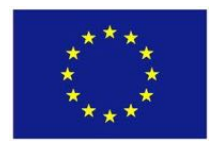

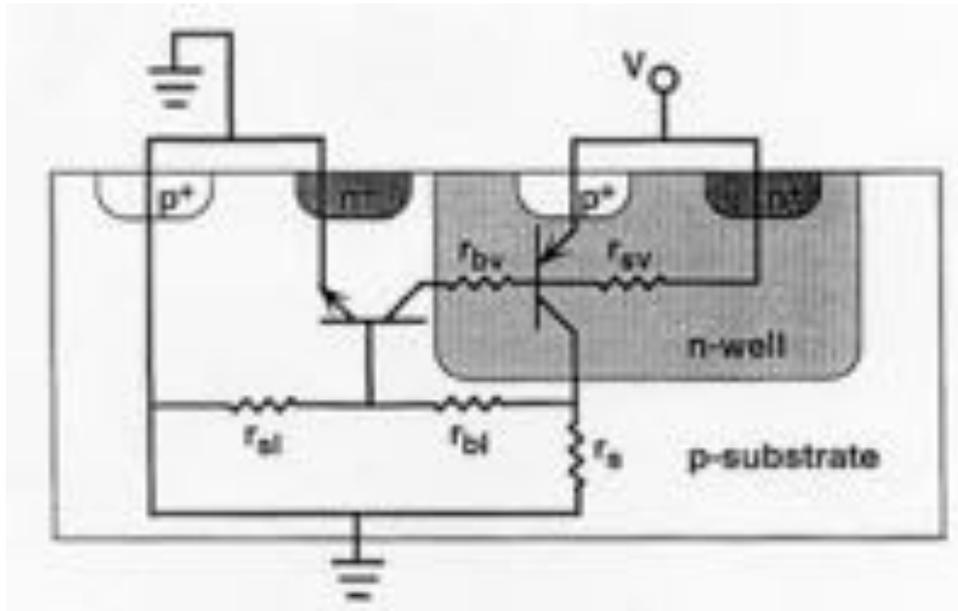

**Figure 1 Simplified two transistor model for Latch-up. Adapted from [RD06].**

Single event transient (SET) are momentary disturbances generated by recoiling ions affecting combinational gates. SET is mainly of concern to clocked digital components. Spurious signals within one clock cycle propagate asynchronously on signal lines, causing latching of erroneous bit values. SETs can propagate through digital design causing seemingly random logic errors. SET is strongly dependent on incident LET value as the amplitude and duration of the spurious signal determine if a latch or a memory element shift in value occurs. The effect can be mitigated by adding nodal capacitance to the clock line, careful clock tree design and introducing clock buffers [RD01].

Total ionizing dose (TID) is the cumulative effect of ionizing radiation.

Gamma rays interact with matter in three different ways:
- Photoelectric effect
- Compton scattering
- Pair production (electron-hole pair)

In silicon, the photoelectric effect dominates at photon energies less than 50 keV, and pair production dominates at energies greater than 20 MeV with Compton scattering dominating in the intervening energy range. The effect of the gamma radiation in electronics is dependent either on the total accumulated dose or the rate of accumulation. For CMOS electronics the dominant effect is the creation of electron-hole pairs inside the silicon causing long term charging (hole accumulation) in the $SiO_2$ (oxide). When charge is built up in the oxide, threshold voltages starts to shift, noise increase and a temperature dependent drain-source leakage current increase can be seen. Creation of pairs in $SiO_2$ require 17 eV/pair produced [RD03] and the yield is strongly dependent on radiation source, field oxide geometry and electric field parameters. In the case of neutron irradiation the main contributor to total dose is the gamma rays from the neutron source and any secondary

This project is funded by the Horizon 2020 Framework Programme of the European Union. Project number 654124.

gamma rays generated by bremsstrahlung or capture in the lead shielding. The alpha particles generated in the $^{10}B(n,\alpha)^7Li$ reaction also contribute to charge trapping by pair production but at a much lower yield.

A second effect in electronics is the creation of new interface states in the $SiO_2$-Si interface due to chemical bonding changes at the interface. This causes the same effects as charge trapping but with less temperature dependence. In addition it can alter the carrier mobility greatly reducing the electronics performance. TID effects are mainly mitigated at a process level by adding epitaxial layers and ensuring correct oxide resistivity. Thinning of oxides greatly reduce the ability to accumulate charge and smaller process nodes are less susceptible to total dose effects as trapped hole density is approximately proportional to the oxide thickness. At deep sub-micron process nodes, an increase in radiation induced leakage current can be expected.

In general the effect of TID in analogue electronics is seen in gain degradation, noise increase, changes in offset voltages and bias currents. For digital electronics the switching speed can be affected and the digital threshold voltages can shift[RD04].

## 1.2   Effects of Neutron Radiation

Neutrons are classified based on energy into:

- Relativistic               >20 MeV ($\lambda$ < 6 fm)
- Fast                       1-20 MeV (6 fm < $\lambda$ < 0.1 fm)
- Intermediate               300 eV-1 MeV (0.1 fm < $\lambda$ < 1.7 pm)
- Resonance                  10-300 eV (0.1 pm < $\lambda$ < 9 pm)
- Slow                       1-10 eV (9 pm < $\lambda$ < 0.3 pm)
- Epicadmium                 0.6-1.0 eV (0.3 pm < $\lambda$ < 0.4 pm)
- Cadmium                    0.4-0.6 eV (0.4 pm < $\lambda$ < 0.5 pm)
- Epithermal neutrons        0.025-0.4 eV (0.5 pm < $\lambda$ < 1.8 Å)
- Thermal neutrons           0.025 eV ($\lambda$ = 1.8 Å)
- Cold neutrons              1e-5-0.025 eV (1.8 pm < $\lambda$ < 90 Å)
- Ultra-cold neutrons        1e-7 eV ($\lambda$ = 900 Å)

Neutrons interact with matter predominantly through strong interaction. Strong interactions take place when neutrons approach the atom nucleus within $10^{-13}$ m. The main effects are listed below and are dependent on neutron energy [RD05].

**Elastic scattering**

The most important mechanism for energy loss in the MeV range.

**Inelastic scattering**

The nucleus is left in an exited state which can lead to decay by gamma ray emission. Inelastic scattering typically occur if the neutron has energy in the range 1 MeV or more (fast and relativistic neutrons).

**Radiative neutron capture**

This project is funded by the Horizon 2020 Framework Programme of the European Union. Project number 654124.

Neutron capture cross section is inversely proportional to the neutron velocity. The likelihood for neutron capture increase with lower energies. In silicon devices boron ($^{10}$B) is present in passivation layers. The isotope $^{10}$B has a large thermal-neutron cross section for the $^{10}$B(n,α)$^{7}$Li reaction. I.e. when subject to neutron irradiation, boron yield alpha particles and lithium ions with LET value 1 MeV/cm2/mg and 3 MeV/cm2/mg respectively with ranges up to 5 μm in silicon.

**Nuclear reactions**

Neutrons are captured and charged particles are emitted. The reactions occur from the eV range to the keV range. The cross-section is diminished at lower energies.

**Fission**

Fission can occur with neutron with energies below 10 keV interact with elements with proton number Z>56.

**High energy hadron showers**

Hadron showers can occur for relativistic neutrons.

## Irradiation Tests

Neutron detection test have been performed at the neutron source at LLB in Saclay, France. The prototype instrument prepared for preliminary tests has been used together with the neutron scintillator [RD13] for those tests.

The two main goals for the test campaign were to characterize the scintillator and evaluate the stability of the prototype with respect to the applied radiation dose from the neutron beam.

Due to the limited availability of the beam facility the number of tests and the number of tested devices have to be limited. It is intended to achieve qualitative measurements which indicate the radiation harness of the prototype which is designed without protection against radiation.

Radiation effects are expected within the analogue and also within the digital parts of the prototype. By first principle it is expected that the analogue part of the prototype will lose gain and therefore the output signal will be weaker with increased total dose. In contrast to that the effects of the digital parts will be more severe and might cause sudden stop of the firmware implemented on the prototype. The communication link to the prototype will break-down completely in such a case.

The obtained results are compared to the theoretically expected values for radiation hardness.

## 1.3   Performed Radiation Hardness Tests

Two different tests are performed to evaluate the radiation hardness of the prototype system. The prototype system is referred as "device under test" (DUT) during the test

This project is funded by the Horizon 2020 Framework Programme of the European Union. Project number 654124.

campaign:

**Test A:** The neutron detection gain of the DUT is measured. The prototype delivers output results with a distinctive output value per detected neutron. Other secondary events/particles will also be measured during operation of the DUT (noise from various sources, gamma-radiation), but those events shall be ignored and only the main peak values originated from neutron detection shall be evaluated for this test.

This test requires the DUT to be powered and fully operational.

Expected outcomes of the measurements:

- Expected output value per detected neutron: The exact value of the output value peak originating from neutron detection is determined during the pre-test of the test campaign. It will also vary from channel to channel due to the properties of the detection principle (scintillator and multi anode photomultiplier, different gain factors for the individual pixels). The value is expected to be in the range of 1/3 of the total dynamic range of the DUT. The total dynamic range is 0pC to -200pC with a resolution of 8 bit.
- Expected change in the performance due to neutron irradiation: With respect to the different irradiation effects described above two main cases can occur. Firstly the DUT might stop working due to irradiation damages/effects, secondly the gain of the DUT could change. In the second case the peak output value during the neutron detection tests will change.

**Test B:** The operational status of the DUT after irradiation with the maximum available neutron radiation dose shall be evaluated. The goal of this test is to determine if the DUT can still be operated after high doses of radiation and if the occurred effects are reversible. The DUT will be switched off when non-functionality is detected and left for annealing since it is expected to be activated after the tests. In a second part of this tests the functionality of the DUT is tested outside of the neutron radiation facility.

This test requires the DUT to be powered and fully operational.

Expected outcomes of the measurements:

- The DUT is expected to stop working after a certain radiation dose. After annealing of the DUT the functionality is evaluated.

## 1.4 Irradiation source

The Laboratoire Leon Brillouin (LLB) at Saclay, France has provided access to the neutron beam from its 14 MW fission reactor. The test setup provides a neutron beam with a wavelength between 1 Å and 14 Å with a luminance peak at ca. 3 Å , see Figure 2. These are cold and thermal neutrons due to the moderation in the reactor.

The neutron flux of the beam is 1e8 neutrons per second per cm$^2$. The beam setup is seen in Figure 3.

This project is funded by the Horizon 2020 Framework Programme of the European Union. Project number 654124.

## LLB Beam Luminance

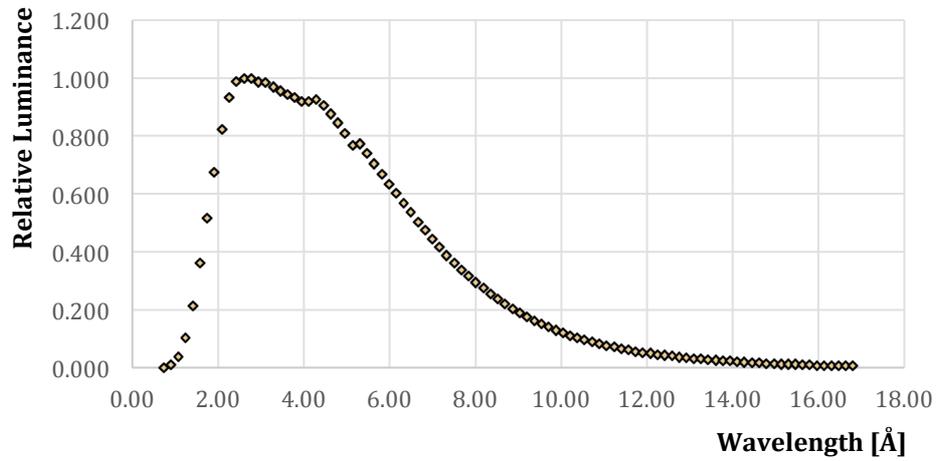

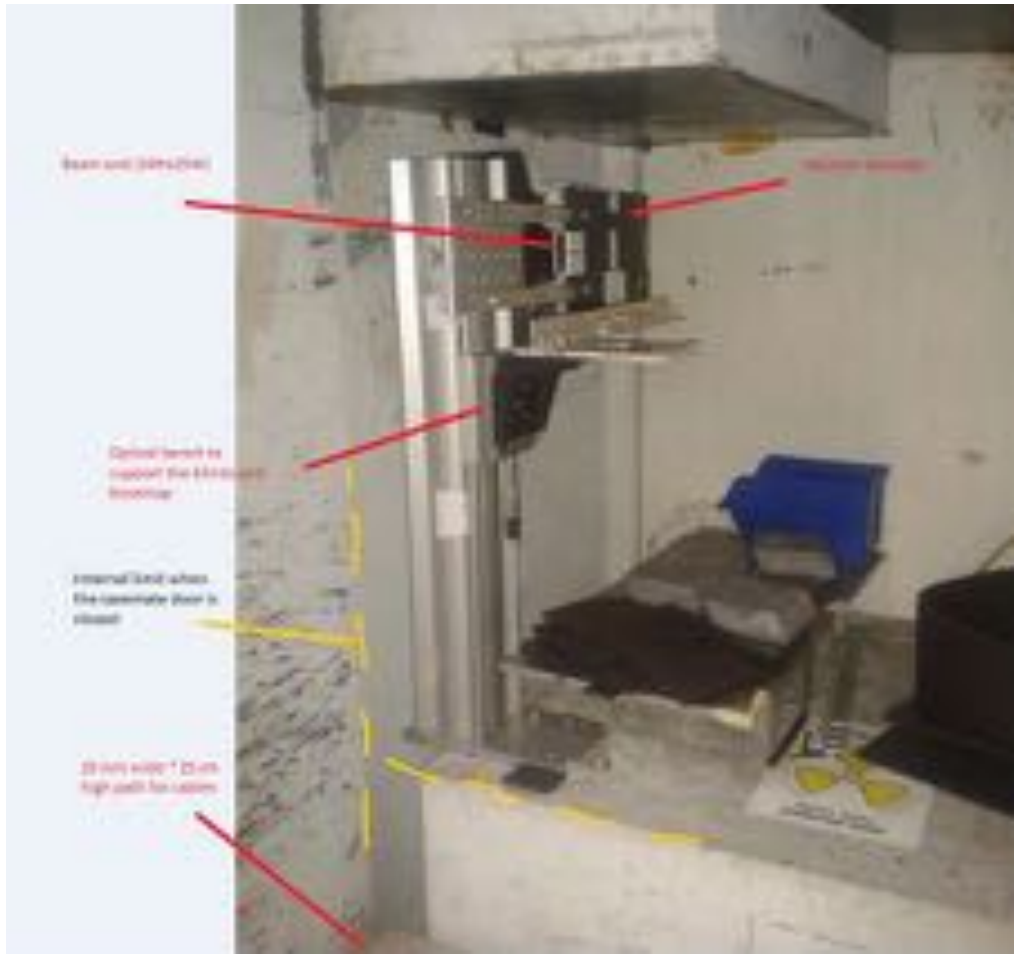

### 1.4.1 Comparison to other neutron sources

The wavelength distribution of the neutrons is close to what is expected from the new European Spallation Source which is built in Lund/Sweden at the time of writing.[RD10]

The total accumulated number of neutrons during the tests per day will be in the range of a several years dose during normal use at comparable neutron sources [RD12].

The results of the irradiation tests will be put into relation to the total dose of neutrons accumulated over the duration of the test campaign.

## 1.5   Device Under Test and Setup

For these tests the prototype system has been used together with the scintillator as shown in Figure 4 [RD13]. The prototype consists of the ROSMAP system delivered by IDEAS and an attached multi anode PMT tube H8500C from Hamamatsu. The design and properties of the scintillator is described in [RD14].

Seen from the beam entrance into the irradiation station, the scintillator and the photomultiplier tube will shield the electronics and therefore reduce the total dose on the electronics alone. This factors have been considered in the conclusion section in chapter 4.

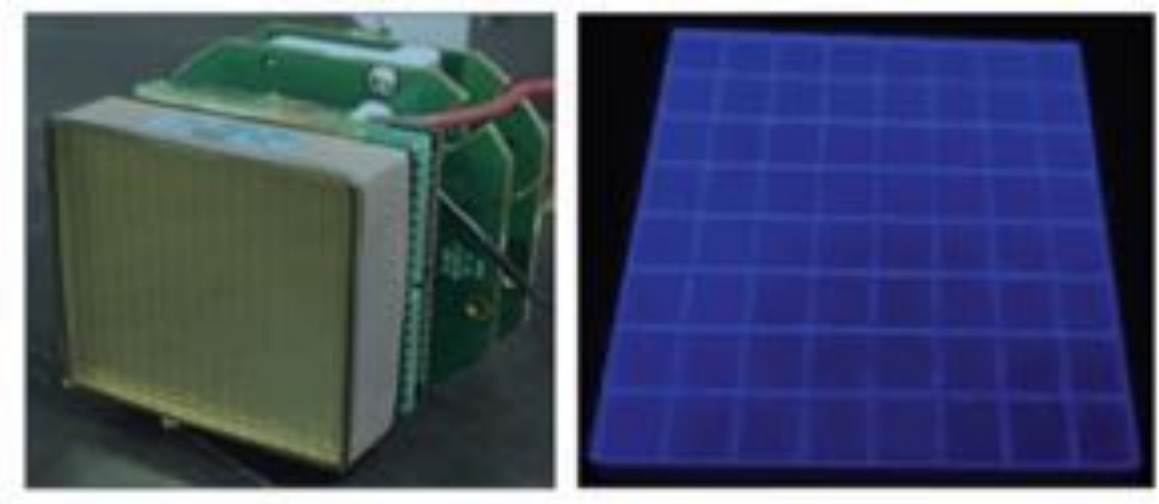

Figure 4 To the left: ROSMAP with mounted H8500C maPMT. To the right: pixelated scintillator.

The ROSMAP system is designed to read-out the used multi anode photomultiplier tube (MaPMT) used for the preliminary tests (64 channel MaPMT type H8500C from Hamamatsu).

The ROSMAP system is self-triggering and delivers output values respective with the peak values of each event detected by the MaPMT/scintillator. The general results from neutron radiation are expected to show low energy events on the scintillator originating from gamma radiation and a distinctive peak originating from neutrons hitting the scintillator. The expected spectrum of events is shown in figure Figure 5.

This project is funded by the Horizon 2020 Framework Programme of the European Union. Project number 654124.

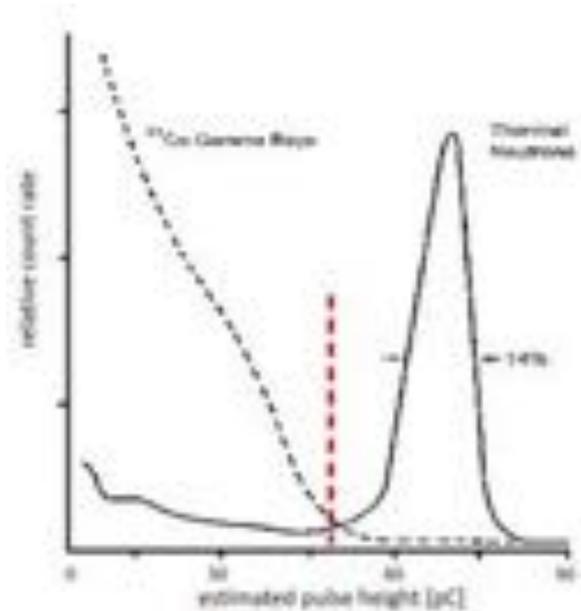

Before the start of the neutron irradiation test campaign, the ROSMAP prototype system functional parameters has been tested in a lab environment. Those will be compared to the functional parameters obtained during a functional test after the main test campaign.

# 2 Results

## 2.1 Test A

The DUT has been irradiated for a total time of ca. 6 days and 11 hours. Several datasets have been acquired during this time. The DUT was fully operational during this time. Downtimes of the beam due to reactor related issues have been already subtracted from the irradiation time.

The datasets acquired show the output values from the prototype during irradiation. Those datasets only include long-term neutron measurements with the DUT. Several tests have been made with the DUT before the long term tests while the neutron beam was switched on and the DUT was powered on.

The result files in Table 1 and Table 2 are used to determine performance and gain changes in the DUT. The beam centered on pixel 27 can been seen in Figure 6. The resulting gain change was found to be 5.154e-13 PeakValue/neutron/cm$^2$, see Figure 7.

This project is funded by the Horizon 2020 Framework Programme of the European Union. Project number 654124.

Table 1 Test parameters for Test A

| Parameter | Value | Notes |
|---|---|---|
| **Rate on the DUT** | 1e8 neutrons per second per cm2<br><br>3.6e11 neutrons per day per cm2 | The illuminated area of 25mm x 25mm by the beam is smaller than the total area of the DUT (ca. 50mm x 50mm) |
| **Total accumulated irradiation time** | 6days 11hours | |
| **Total accumulated neutron dose** | 4.95 e13 neutrons per cm2 | |

Table 2 Result files with respective total dose

| Data File | File Date and Time | Accumulated Dose [neutrons per cm2] [1] |
|---|---|---|
| **049-ROSMAP_daq_2015-10-20_10h-53min-28s.dat** | 20.10.15 10:53 | 3.132e+13 |
| **050-ROSMAP_daq_2015-10-20_11h-33min-07s.dat** | 20.10.15 11:33 | 3.15e+13 |
| **051-ROSMAP_daq_2015-10-21_08h-24min-45s.dat** | 21.10.15 08:24 | 3.942e+13 |
| **052-ROSMAP_daq_2015-10-21_10h-36min-25s.dat** | 21.10.15 10:36 | 4.014e+13 |
| **053-ROSMAP_daq_2015-10-22_10h-18min-09s.dat** | 22.10.15 10:18 | 4.878e+13 |
| **054-ROSMAP_daq_2015-10-22_11h-54min-54s.dat** | 22.10.15 11:54 | 4.95e+13 |

---

[1] Dose accumulated between start of irradiation until the start of the measurement.

This project is funded by the Horizon 2020 Framework Programme of the European Union. Project number 654124.

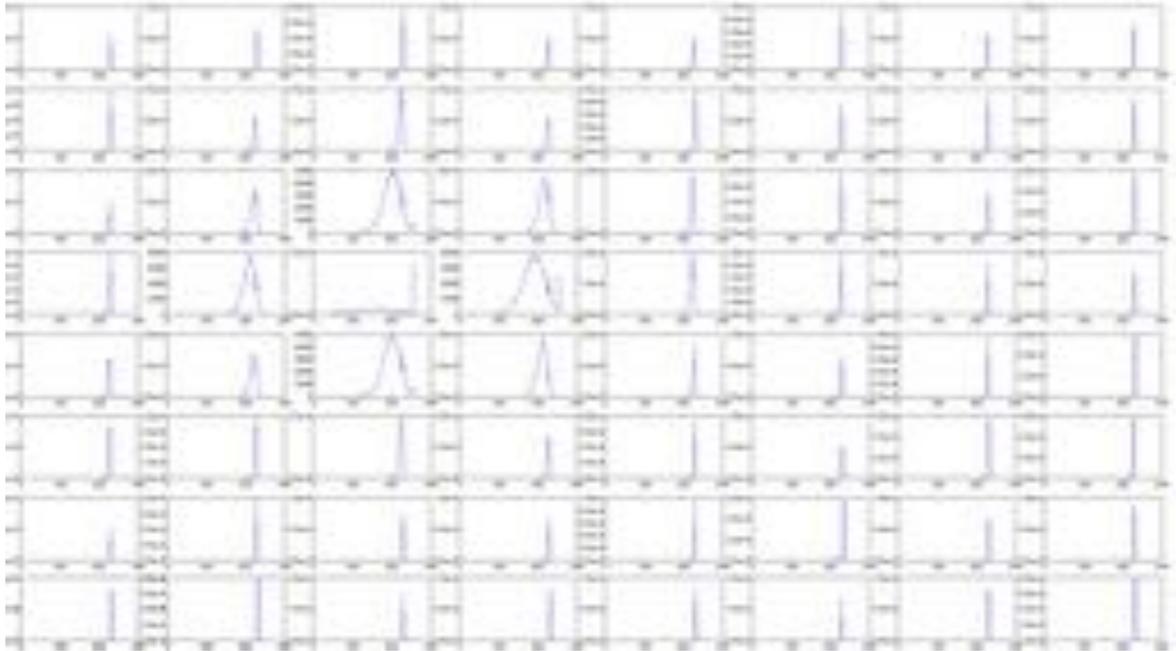

**Figure 6 Histogram per pixel during run 50 show the beam centered on pixel 27, with activity as expected in neighboring channels.**

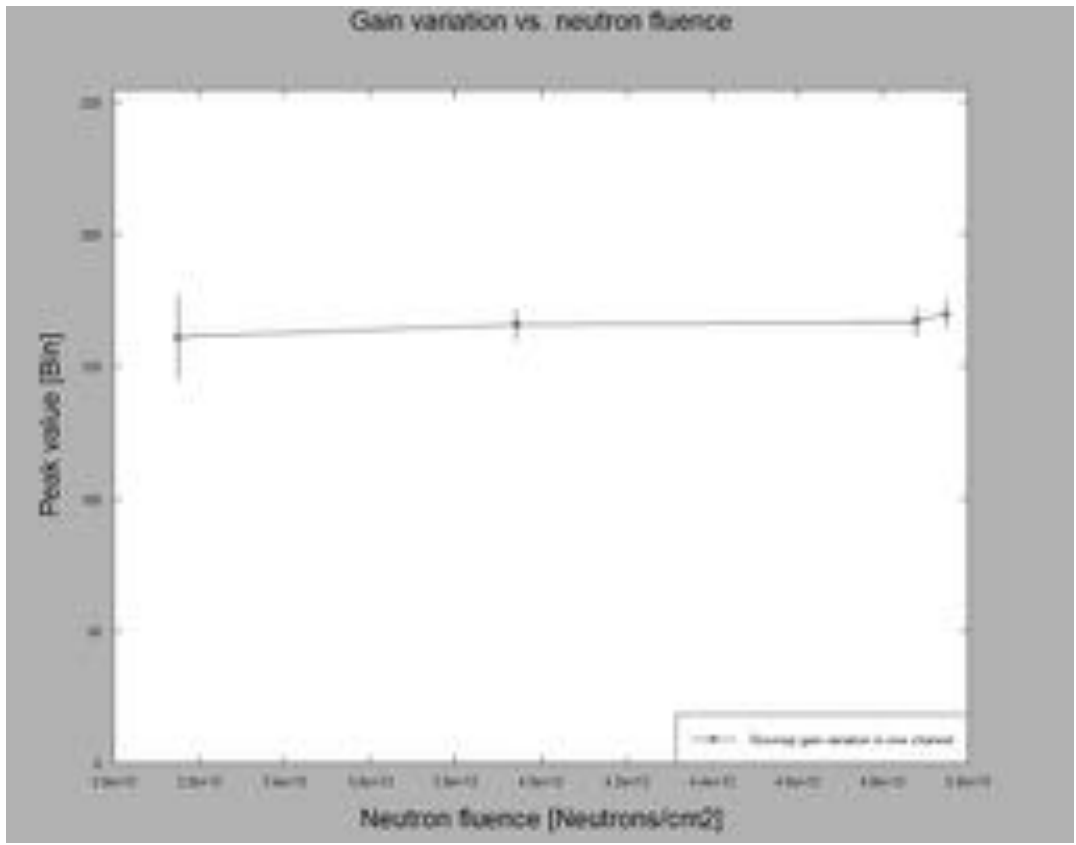

**Figure 7 Gain variation in pixel 27 (center of beam) as a function of fluence for a period of three days (run 49-53). The result values are negative with a bias of an ADC value of 230. Therefore the max value drops slightly with increasing neutron fluence.**

This project is funded by the Horizon 2020 Framework Programme of the European Union. Project number 654124.

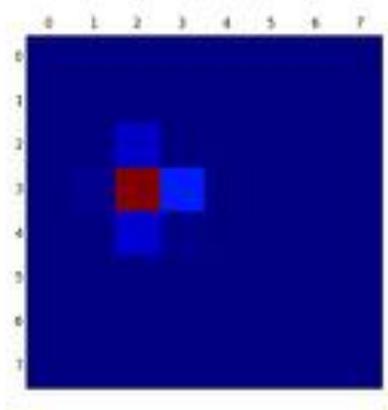

Figure 8. Integrated image of run 51, only the triggering pixels are added up. The peak position at pixel 27 is very well visible.

Neutron detection results:

After the dose of ca. $5\times10^{13}$ neutrons the communication line to the DUT broke down and the DUT was no longer operational.

Two observations have been made towards the end of Test A:

The gain of the DUT drops slightly towards the end of the test / failure of the DUT.

After failure of the DUT and therefore at the start of Test B it has been observed that the DUT failed to start on several occasions. At those occasions the supply current to the DUT was lower than expected (ca. 270mA vs. the expected ca. 410mA). This decreased supply current fits with the expected current consumption of the DUT during programming of the on-board FPGA (start-up procedure). These issues with the startup process were also observed before the irradiation runs. As this behavior could not be reproduced we can only speculate as to the reasons. One possible explanation is a timing error between program startup, computer startup and plugging in of the DUT. Performing all these tasks too fast always resulted in a communication error, however we were not able to find a suitable waiting time to prevent these issues. Later versions will have a more robust timing/start up procedure.

## 2.2 Test B

In addition to the total neutron irradiation dose accumulated during Test A the DUT has been irradiated furthermore in a powered state to evaluate possible reversible/irreversible causes of failure.

This project is funded by the Horizon 2020 Framework Programme of the European Union. Project number 654124.

| Parameter | Value | Notes |
|---|---|---|
| **Rate on the DUT** | 1e8 neutrons per second per cm2<br><br>3.6e11 neutrons per day per cm2 | The illuminated area of 25mm x 25mm by the beam is smaller than the total area of the DUT (ca. 50mm x 50mm) |
| **Total accumulated irradiation time at start of Test B** | 6days 11hours | |
| **Total accumulated irradiation time at end of Test B** | 13days 11hours | |
| **Total accumulated neutron dose** | 1.16 e14 neutrons | |

Total neutron irradiation dose results:

After the applied total dose of ca. $5 \times 10^{13}$ neutrons during Test A the DUT the previously observed startup issues did not allow further operation of the module. The total dose has been increased during Test B to over $1 \times 10^{14}$ neutrons. After demounting of the test setup the DUT has been left for annealing. After annealing, complete removing and redoing all the cabling the device was functional again and was checked in a dedicated testing setup at an IDEAS laboratory.

The DUT has been tested for functional parameters and compared to the functional test from before the test campaign, see Table 4.

This project is funded by the Horizon 2020 Framework Programme of the European Union. Project number 654124.

**Table 4 Functional parameters before and after the irradiation test campaign**

| Test # | Description | Result before irradiation | Result after irradiation |
|--------|-------------|---------------------------|--------------------------|
| T1 | Visual inspection: | OK | OK |
| T2 | Mechanical mating: | OK | OK |
| T3 | Labeling: | 3-6 | 3-6 |
| T4 | Validity of firmware: | DevID: 7016 FW: v1.2.0 | DevID: 7016 FW: v1.2.0 |
| T5 | Communication speed: | 19200 | 921600 |
| T6 | Communication address: | 06 | 08 |
| T7 | Communication settings confirmed: | OK | OK |
| T8 | Set Mbias to 700uA (V1/V2): | OK (-231mV, -265mV) | OK (-243mV, -275mV) |
| T9 | Bias settings confirmed: | OK | OK |
| T10 | Set trigger threshold value (Readback): | 1260mV (1257mV) | 1377mV (1374mV) |
| T11 | Output linearity of 4 selected channels: | N/A | N/A |
| T12 | Functionality of all channels: | OK | OK |
| T13 | Current consumption at 5.0V: | 0.41A | 0.407A |
| T14 | Dynode DC value: | 1204mV | 1204mV |

This project is funded by the Horizon 2020 Framework Programme of the European Union. Project number 654124.

The most important functional parameters with respect to these tests are:

- The generating voltage of the main current bias of the system (T8): A change here would indicate irradiation damage in the analogue part of the ASIC. The slight deviation observed here is within acceptable limits and does not indicate irradiation damage.
- Set trigger threshold (T10): An increased threshold level would indicate a higher noise in the system due to irradiation damage in the analogue front end of the system. The increased value from 1260mV to 1377mV originates from an additional cable attached to the trigger input of the system. This has been done during the preliminary test session to inspect the trigger input of the system prior to the neutron irradiation test campaign.
- Current consumption (T13): A change in power consumption of the module would indicate that parts of the system are damaged due to irradiation. This value is also within acceptable limits and does not indicate irradiation damage.

All parameters of the DUT are back to normal after the test campaign and annealing of the DUT. Therefore all irradiation effects on the DUT were found to be temporary.

# 3  Materials Analysis

To evaluate the absorption rates of the different parts of the prototype a neutron-radiography is taken with and without attached multi anode photomultiplier tube. This is used to determine the total neutron radiation dose applied to the electronical components of the prototype. It is expected that the photomultiplier tubes absorb a certain percentage of the neutron radiation.

## 3.1  Absorption of the Prototype (ROSMAP)

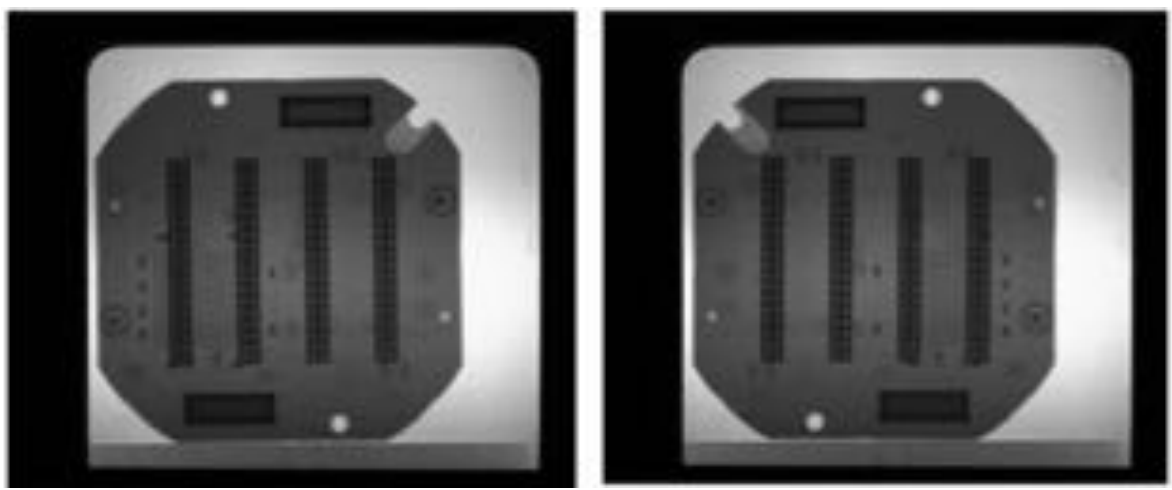

Figure 9 Neutron radiography of the ROSMAP. Radiographies taken from both sides of the prototype.

This project is funded by the Horizon 2020 Framework Programme of the European Union. Project number 654124.

The radiography in Figure 9 shows the parts of the ROSMAP system, which absorbs most of the neutrons. In the pictures the semiconductor components cannot be seen, therefore they barely absorb any neutrons. For the most part only plastic-like materials absorb/scatter neutrons, including connectors, the circuit board material and to a little extend the plastic package of some of the electronic components. This is due to the high hydrogen content in the used plastics.

As a comparison visible light photographs of the ROSMAP system are shown in Figure 10.

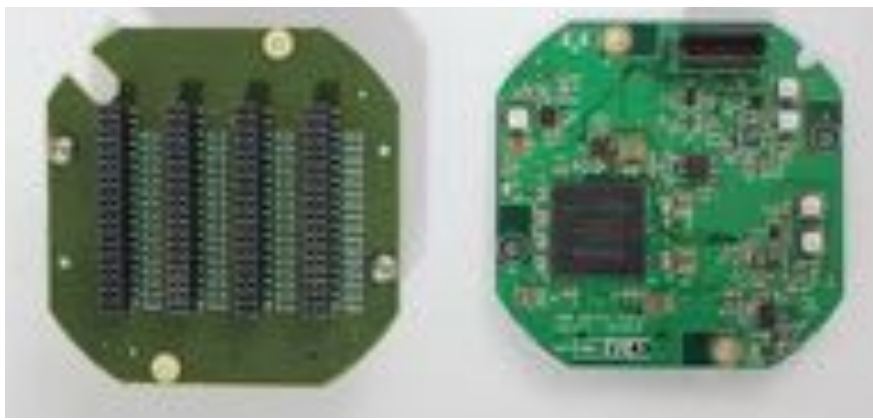

Figure 10 Photograph of the ROSMAP (top and bottom side).

## 3.2    Absorption of the MaPMT

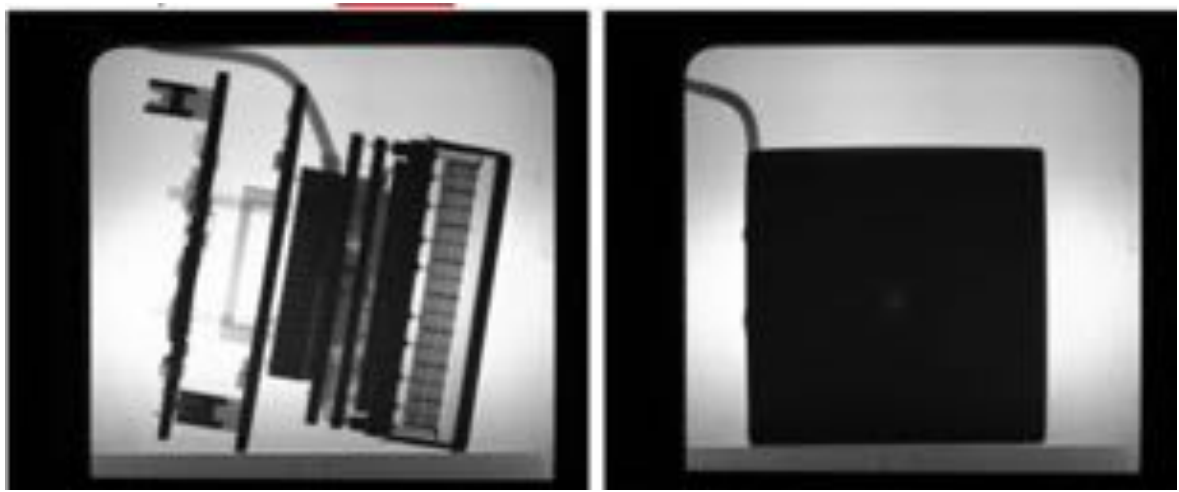

Figure 11 Neutron radiographies taken from the ROSMAP system with the attached H8500c tube from Hamamatsu (left) and the Hamamatsu H8500c tube taken from the front. The slightly lighter point in the center is the mounting point of the module.

Figure 11 shows a high absorption of the photo multiplier tube. Image processing of the radiographies show that the front glass of the used H8500c photomultiplier has an absorption rate of 93%. Therefore only 7% percent of the neutrons reach the electronics

This project is funded by the Horizon 2020 Framework Programme of the European Union. Project number 654124.

components behind the photomultiplier tube. These values were found by comparison of the grey values of the images.

## 3.3  Absorption of the Scintillator

The absorption rate of the used scintillator as described in [RD14] is about 75% to 80%.

# 4  Conclusions for further Development

## 4.1  Functionality of the Prototype (ROSMAP)

The neutron irradiation test campaign showed which dose of neutron irradiation the used prototype system is able to withstand. The used prototype is functional up to a dose of $5 \times 10^{13}$ neutrons in the neutron beam. Up to a dose of $1 \times 10^{14}$ neutrons no permanent damage was found.

For the above dose results to hold true the absorption rates of the scintillator and photomultiplier tube have to be taken into account. Those two components absorb ca. 75% and 93% respectively of the neutron radiation. Therefore the electronics is only irradiated by a dose corresponding to ca. 5.5% of the total dose. The final conclusions for the used prototype are:

Table 5 Conclusion on maximum neutron dose. The absorptions of the materials shielding the electronics are calculated into the determined values.

| Conclusion | Determined Value |
|---|---|
| **Maximum neutron irradiation dose for functional prototype** | $2.75 \times 10^{11}$ neutrons/cm$^2$ |
| **Maximum neutron irradiation dose without permanent damage** | $5.5 \times 10^{11}$ neutrons/cm$^2$ |

No SEE or TID related effects were detected during operation. The analogue channel gain change in response to the radiation is 5.154e-13 bins/neutron/cm$^2$.

Comparative studies performed by JAXA and NASA of IDEAS ASICs implemented using the standard 0.35 µm library have shown TID tolerance up 20 Mrad [RD07] and SEL LET$_{th}$ at 8 MeV/cm$^2$/mg [RD08]. The received gamma dose in the test will not affect the ASIC, neither will any ionizing particles with LET > 3 MeV/cm$^2$/mg occur. Hence no SEE in the ASIC is expected to occur due to the ESS neutron radiation.

This project is funded by the Horizon 2020 Framework Programme of the European Union. Project number 654124.

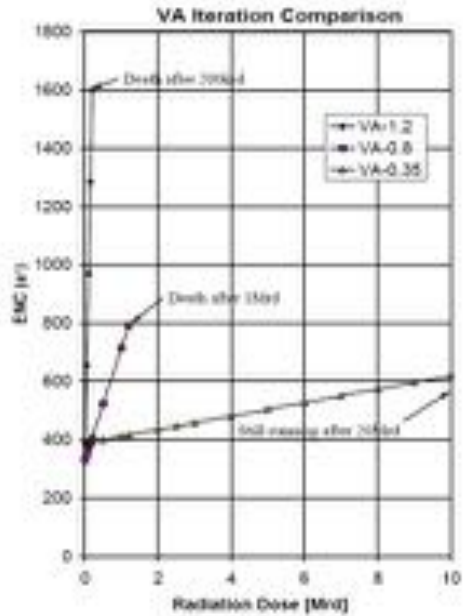

Figure 12 Equivalent noise charge (ENC) versus radiation dose, from [RD07].

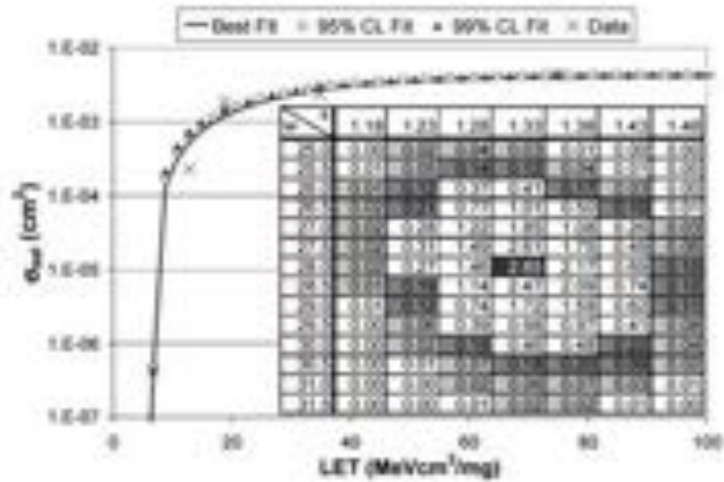

Figure 13 Prior for a Weibull fit to a SEL CS vs LET curve for the XA1 ASIC on the Swift BAT telescope, from [RD08].

The Xilinx Spartan-6 (XC6SLX16) is implemented in a boron free 45 nm process and hence the $^{10}B(n,\alpha)^7Li$ reaction cannot occur inside the FPGA in response to impinging neutrons. The package is a low alpha emitting package reducing the risk of SEE. Quinn et al. [RD11] report results that indicate that both the block RAM and the configuration memory are sensitive to SEU and that the FPGA is sensitive to SEFI. TID results of the Spartan-6 indicate that the 40 nm node is operational to 380 krad and that the SEE LET threshold is below 1 MeV/cm$^2$/mg [RD09]. Xilinx state that the FPGA can operate in neutron energies up 10 MeV.

This project is funded by the Horizon 2020 Framework Programme of the European Union. Project number 654124.

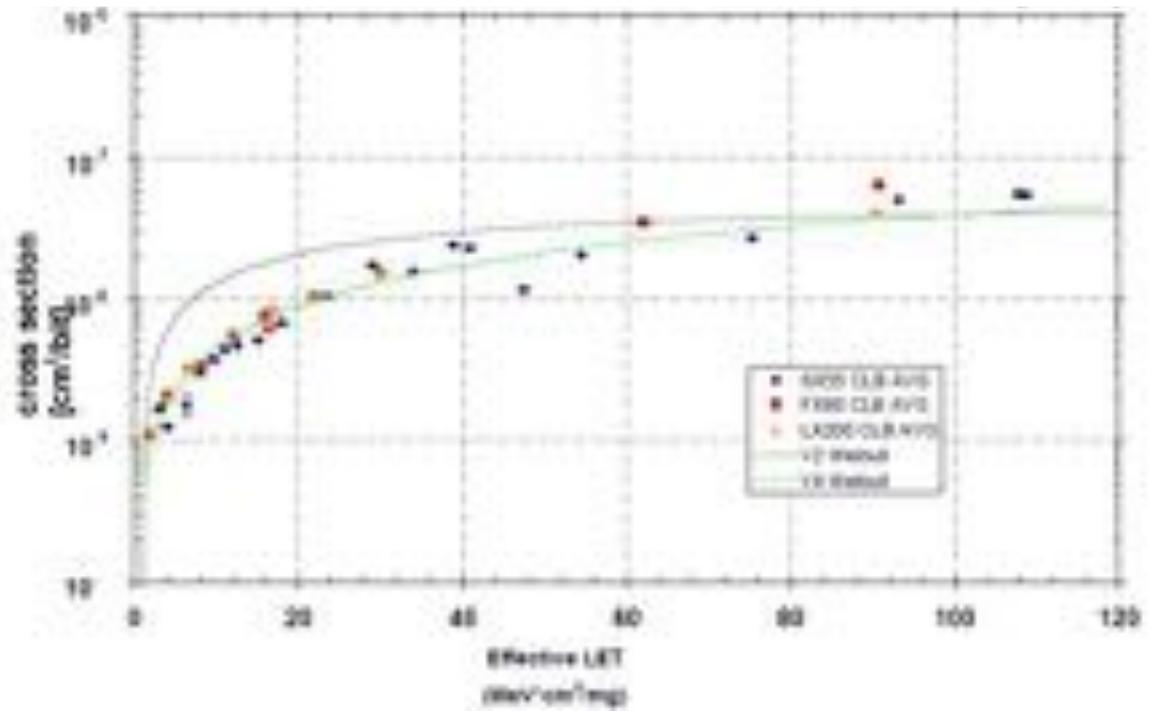

## 4.2   Required Shielding

The required shielding is dependent of the used facility, the wavelength spectrum and the rate of the neutron beam. Furthermore changes in the absorption factors of the used scintillator and possibly of changes in the composition of the photomultiplier tube have to be taken into account.

The conclusions in chapter 4.1 have to be taken into account in case the electronic components are not shielded at all.

This project is funded by the Horizon 2020 Framework Programme of the European Union. Project number 654124.

This project is funded by the Horizon 2020 Framework Programme of the European Union. Project number 654124.

# SoNDe

## Solid-State Neutron Detector

### INFRADEV-1-2014/H2020

### Grant Agreement Number: 654124

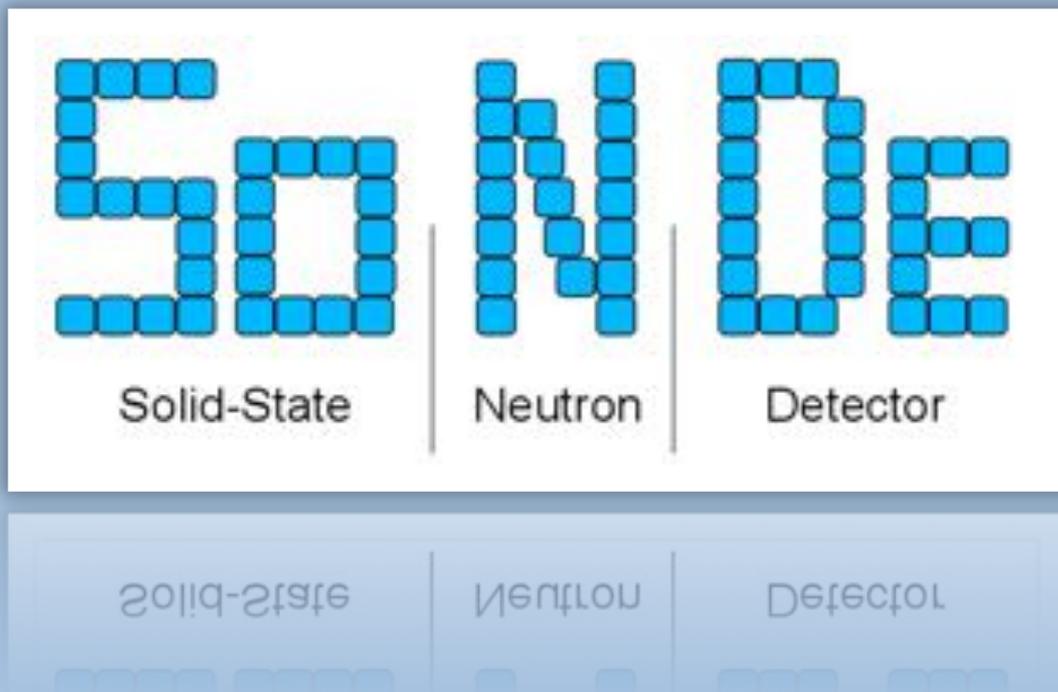

Deliverable Report: D5.1 Additional Applications Report

**This project is funded by the Horizon 2020 Framework Programme of the European Union. Project number 654124.**

## Project and Deliverable Information Sheet

| SoNDe Project | Project Ref. No. 654124 | |
|---|---|---|
| | Project Title: Solid-State Neutron Detector SoNDe | |
| | Project Website: http://www.fz-juelich.de/ics/ics-1/DE/Leistungen/ESS/SoNDe-Projekt/ | |
| | Deliverable ID: D5.1 | |
| | Deliverable Nature: Potential applications | |
| | Deliverable Level: PU | Contractual Date of Delivery: 12.2015 |
| | | Actual Date of Delivery: 03.2016 |
| | EC Project Officer: Bernhard Fabianek | |

## Document Control Sheet

| Document | Title: Solid-State Neutron Detector SoNDe | |
|---|---|---|
| | ID: AdditionalApplicationsReport-D5.1 | |
| | Version: 1.0 | |
| | Available at: http://www.fz-juelich.de/ics/ics-1/DE/Leistungen/ESS/SoNDe-Projekt/Reports/_node.html | |
| | Software Tool: MS Word 2011 | |
| | Files: Additional potential applications –D5.1.docx | |
| Authorship | Written by | Sylvain Désert, LLB |
| | Contributors | Frederic Ott, LLB Sebastian Jaksch, FZJ |
| | Reviewed by | Sebastian Jaksch, FZJ |
| | Approved | Sebastian Jaksch, FZJ |

# List of Abbreviations

| | |
|---|---|
| FZJ | Forschungszentrum Jülich, Jülich Research Centre |
| JCNS | Jülich Centre for Neutron Science |
| LLB | Laboratoire Léon-Brillouin |
| ESS | European Spallation Source |
| CEA | French Agency for Atomic Energy |

# List of Figures

**This project is funded by the Horizon 2020 Framework Programme of the European Union. Project number 654124.**

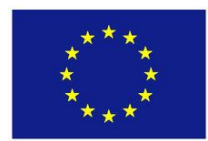

# Table of Contents

This project is funded by the Horizon 2020 Framework Programme of the European Union. Project number 654124.

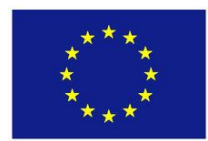

# Executive Summary

This report lists the additional potential applications using the SoNDe project technology, initially build to provide suitable neutron detectors for existing sources but also for the new high brilliance source as ESS, and presents the strategy adopted within CEA to communicate the project to industrials hence keeping new applications opened up for the future.

# Introduction

Within the SoNDe-project a neutron detector based on a solid-state scintillator is to be developed. The technology used within the SoNDe project will allow manufacturing detectors able to cope with extremely high count rates of about 20 MHz at 10% dead time on a 1 m² detector area. Such high counting rate will be compulsory with the use of the next generation of high brilliance neutron sources such as ESS. The detectors will be composed of units of 5x5 cm² which can be assembled at will to produce any geometry. The use of an optimized scintillator will allow an efficiency of 80% in the range 0.2 to 2 nm (0.2 to 20 meV). Also the scintillator may be changed to fit other incoming particles or wavelengths. Apart from its initial purpose for neutron science, all of the above listed key technologies can be useful to various applications hereafter documented.

## Positron Emission Tomography

Positron emission tomography (PET)[1] is a nuclear medicine, functional imaging technique that produces a three-dimensional image of functional processes in the body. The system detects pairs of gamma rays emitted indirectly by a positron-emitting radionuclide (tracer), which is introduced into the body on a biologically active molecule. Three-dimensional images of tracer concentration within the body are then constructed by computer analysis.

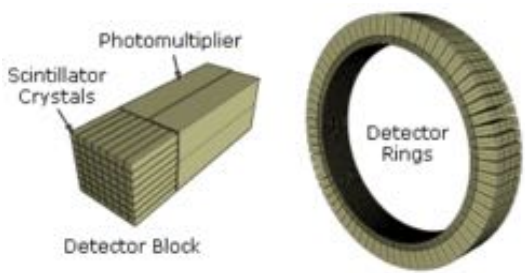

Fig 1. Schematic view of a detector block and ring of a PET scanner

The detector required for PET is very similar to our technology since they both use a scintillator to convert the incoming radiation (neutron or positron) into the visible range. Typical spatial resolution is of the order of 4 mm[2]. Getting to smaller pixel size would result in too many cristals and electronics to handle. The advantage using the SoNDe would thus be its straight possibility to provide pixel sizes of 5 mm, still achieving sub resolution down to 1 mm with the Anger technique and being compatible with a use in harsh environment due to its already proven robustness to radiation (see deliverable 1.3 *Report on radiation hardness of electronics*). Finally, the detector ring geometry can be achieved using appropriate geometry for the mechanics of the detector.

This project is funded by the Horizon 2020 Framework Programme of the European Union. Project number 654124.

## Non-destructive testing and analysis

### Building industry

Neutron radiography is a non-destructive technique, eventually complimentary to X-ray radiography as the contrast elements are different[3] allowing inspection of volumic components either in-line during process or after production to verify the presence and connection of hidden compounds, find cracks, voids or other flaws in assembled products, verify weld integrity, guide a robot for assembly, count the number of welds, teeth, slots or other such features, perform a critical dimensional measurement... Among examples which can satisfy with mm resolution are the inspections of oil and gas weld, turbine blades, ship assemblies, electronic boards…

### Food industry

The non-destructive analysis can also be used in the large market of food area through in-line analysis during food production as it allows direct impact measurement of the upstream process equipment on the quality of the product, such as moisture content, volumic homogeneity … Such volumic measurements cannot be achieved with optical elements and could be operated without human interference. Among the key processes equipment that could be monitored are the *extrusion process* where optimization of the thermal and mechanical energy input are required to optimize costs, the *drying process* via measurement of the moisture for a nutritient mass balance and also production cost point of view and the *coating* or how uniform has the lipid been absorbed on the product to get nice visual appearance and avoid any contamination.

### Extrusion Production

One challenge in high-throughput extrusion production is the in-line, non-material analysis with an adequate speed to ensure product quality. Today such systems usually use methods such as ultrasound, temperature or pressure as a probe to investigate the extruded material.[4,5] Using a single SoNDe module and a weak radioactive laboratory source these systems could also be run using radiography analysis of the material, which significantly widens the set of parameters that can be monitored during production, thus leading to more reliable products and better control over the production plant.

## Homeland security

Container security[6] is a major task nowadays since approximatively 95% of the world's trade moves by containers, primarily on large ships, but also on trains, trucks, and barges. However, the rise of terrorism and the possibility that a container could be used to transport vehicle for weapons of mass destruction or high explosives have imposed that the security of the shipping container system must be greatly improved. Aside from the direct effects of an attack, the economic, social, and political consequences of a significant interruption in the transport chain would be critical. As a results, International Initiatives from the G8 Summit (2002) and US (Megaport Initiative, 2003) have arisen. Among the defensive tools is the implementation of detectors at communication choke points, with a goal of 100 ports equipped by 2016. Such detectors or Radiation portal Monitor, could be relatively large and sophisticated (Figure 2) and do not require to have high resolution.

This project is funded by the Horizon 2020 Framework Programme of the European Union. Project number 654124.

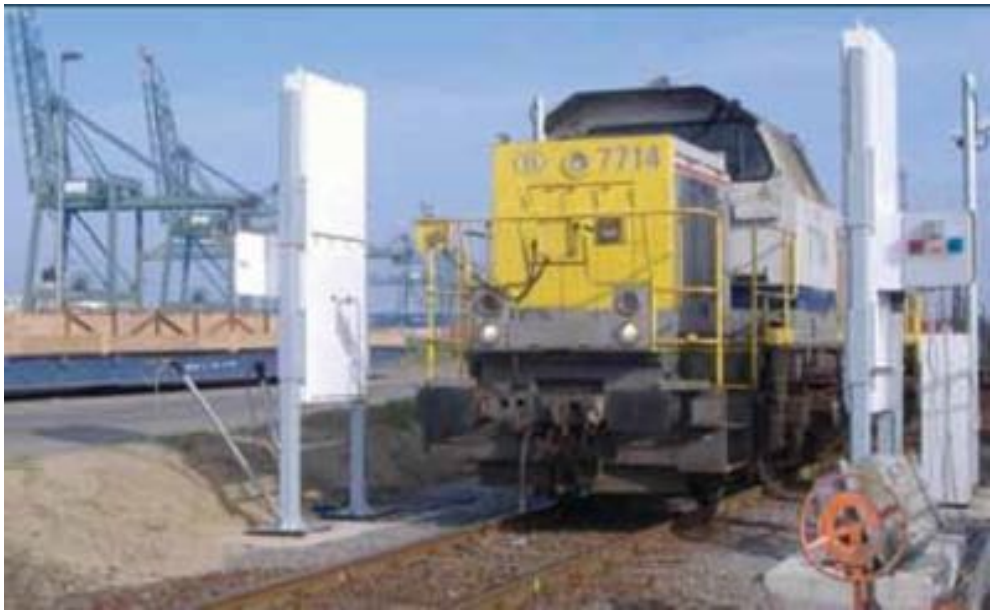

Fig 2. Radiation Portal Monitor (Port of Antwerp, Belgium)[7]

## Oil and gas well logging

Well logging is a mapping technique for exploring the subsurface and a mean to evaluate the hydrocarbon production potential of a reservoir. It is thus of huge concern for the petroleum industry to optimize the process by means of on-line characterization techniques. Specifically, neutron detectors are used to determine the porosity, density and water content in the range 150 to 175°C[8]. Typical detectors are using $^3$He and thus are directly affected by the market price increase of $^3$He. Also, the necessity to harvest deeper into the soil, hence at higher pressure and higher temperature – up to 400°C - pushes the actual technologies to their limit[9] and will require new detectors technologies in a close future. Studies could be performed to test if an embedded SoNDe technology could fulfill such new requirements.

## On-going prospective for new applications

### Technology transfer

The CEA technology transfer office has been contacted. As a result of discussions, first the SoNDe technology has been integrated in the list of CEA technologies so that the technology office can refer to it when answering industrial demands. Also, a technical sheet describing the SoNDe detector features (Figure 3) has been edited and placed on the CEA platform. This media platform is opened in free access online[10] to the public mainly dedicated to the industrials willing to use or collaborate with patented technologies involved within the CEA.

### Workshop

A workshop dedicated to the SoNDe project shall be organized as deliverable 5.5 *Workshop for potential users*. Using the address books of all the partner institutions, this Workshop will be extensively advertised and may reveal new applications.

This project is funded by the Horizon 2020 Framework Programme of the European Union. Project number 654124.

Combined Technology transfer survey and workshop could provide new sources of applications during or after completion of the SoNDe project.

Fig 3. SoNDe technical sheet on the CEA innovation platform

## Conclusion

This document presented the possible applications of the SoNDe detector and showed that they cover a wide range of topics, from medicine to personal safety and industrial inspection and/or process control. Each of these topics represents a high potential demand of SoNDe type detectors. In parallel, a descriptive technological sheet has been published online with free access keeping potential new applications to be found in the future.

In order to get both the attention of potential users and screen for additional application a SoNDe Application workshop is scheduled to take place in October 2016, where both the technology will be presented to potential users as well as their input will be collected for further developments.

**This project is funded by the Horizon 2020 Framework Programme of the European Union. Project number 654124.**

# SoNDe

## Solid-State Neutron Detector

### INFRADEV-1-2014/H2020

### Grant Agreement Number: 654124

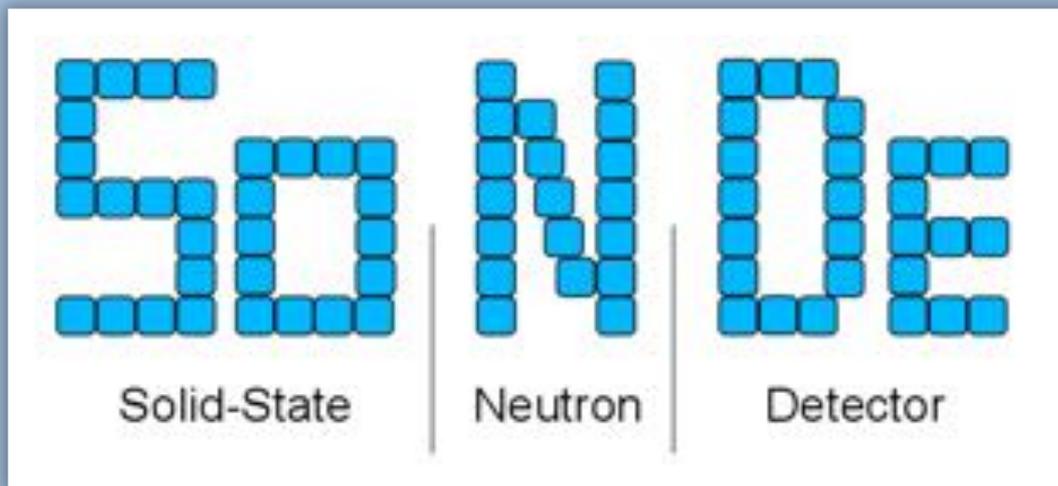

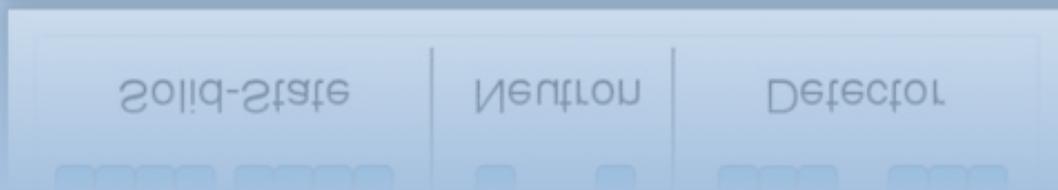

Deliverable Report: D2.1 2x2 Demonstrator

**This project is funded by the Horizon 2020 Framework Programme of the European Union. Project number 654124.**

## Project and Deliverable Information Sheet

| SoNDe Project | Project Ref. No. 654124 | |
|---|---|---|
| | Project Title: Solid-State Neutron Detector SoNDe | |
| | Project Website: http://www.fz-juelich.de/ics/ics-1/DE/Leistungen/ESS/SoNDe-Projekt/ | |
| | Deliverable ID: D2.1 | |
| | Deliverable Nature: 2x2 Hardware Demonstrator | |
| | Deliverable Level: PU | Contractual Date of Delivery: 30.05.2016 |
| | | Actual Date of Delivery: 16.08.2016 |
| | EC Project Officer: Bernhard Fabianek | |

## Document Control Sheet

| Document | Title: Solid-State Neutron Detector SoNDe | |
|---|---|---|
| | ID: Website-Deliverable-D2.1 | |
| | Version: 1.0 | |
| | Available at: Project Website | |
| | Software Tool: MS Word 2011 | |
| | Files: 2x2-Demo-Deliverable-D2.1.docx | |
| Authorship | Written by | Sebastian Jaksch, FZJ |
| | Contributors | Ralf Engels, Günter Kemmerling (FZJ), Codin Gheorge (IDEAS) |
| | Reviewed by | Sebastian Jaksch, FZJ |
| | Approved | Sebastian Jaksch, FZJ |

This project is funded by the Horizon 2020 Framework Programme of the European Union. Project number 654124.

# List of Abbreviations

FZJ            Forschungszentrum Jülich, Jülich Research Centre

JCNS           Jülich Centre for Neutron Science

LLB            Laboratoire Léon-Brillouin

ESS            European Spallation Source

IDEAS          Integrated Detector Electronics AS

MaPMT          Multi-anode Photomultiplier Tube

S-DAM          SoNDe Data Acquisition Module

ROSMAP         IDEAS code name for counting electronics

# Table of Contents

**This project is funded by the Horizon 2020 Framework Programme of the European Union. Project number 654124.**

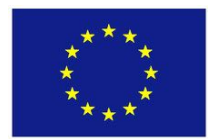

## Executive Summary

This 2x2 demonstrator of a SoNDe detector module shows a combination of four Multi-anode Photomultiplier Tube (MaPMT) with the accompanying electronics similar to the setup projected to be used in the final detector.

It both serves to show the feasibility of the approach to combine several panels of MaPMTs together as well as for testing of the technology. At this stage, a wide array of questions concerning the interoperation of several modules have to be answered, such as the mode of time synchronization between the modules and the combination of data coming from several modules. These tests are fundamental for a proper operation of the complete SoNDe detector at a later stage, however are so deeply ingrained into the construction that any design changes at a later stage will prove very costly, in time and resources. Therefore, a detailed testing of the 2x2 demonstrator is now scheduled and will be performed in the near future.

The demonstrator consists of four identical modules, each with a pixelated scintillator/carrier glass sandwich, a H8500 Hamamatsu MaPMT and a specifically designed readout system with a front-end board to connect to the MaPMT and a controller board that also houses the logic and communication part.

## Introduction

One of the main goals in the first work package WP2 (2x2 Demonstrator) is to determine the scalability of the proposed technology. To do so the interconnection between the modules, the synchronicity of the data transfer and the reassembly of an image from several modules have to be tested and evaluate. Solving these issues at this stage in the project will ensure a working full-scale detector by the end of the project. This testing can only be done if appropriate hardware is available, which is why this 2x2 demonstrator was constructed.

Solutions to other issues that were identified earlier during the project, such as the separation of the single pixels in the scintillator glass, connection between readout electronics and MaPMT and reliability of the software framework can also continuously be studied using this new improved demonstrator.

The main challenges during the development phase of the 2x2 demonstrator were the high integration density of electronics, keeping the form factor of the module below 5x5 cm$^2$ and thus smaller than the MaPMT outer dimension. The overall power consumption of the modules, the electronic and mechanical connections of the single components, that both grants a reliable construction procedure but keeps the modularity and thus ease of maintenance after the modules have been assembled into the whole detector. The 2x2 demonstrator design is in line with all the requirements to the SoNDe detector and we are confident that further testing will prove its feasibility.

## Construction of the 2x2 demonstrator

As previously described the 2x2 demonstrator consists of four pixelated scintillator/carrier glass sandwiches, H8500 Hamamatsu MaPMTs and two-stage readout electronics comprised of a controller board for the logic/communication and a frontend board to provide connections to the MaPMT. All components mounted onto a 2x2 frame for later testing and

This project is funded by the Horizon 2020 Framework Programme of the European Union. Project number 654124.

ease of handling. In the following section, we will describe the construction of the 2x2 demonstrator, which will in itself be the prototype for the later upscaling in WP3.

## Improvements on Pixelated Scintillator/Carrier glass sandwich

In our former deliverable, we were putting forth the approach to separate the single pixels using a wafer saw and physically cutting the scintillation glass that is itself glued to a carrier glass. Further examination of the issue has shown several other approaches, that will allow mounting on an aluminum sheet, which improves mechanical stability considerably. These improvements will be tested using the 2x2 demonstrator setup.

## H8500 Hamamatsu MaPMT

The H8500 MaPMT was purchased from Hamamatsu. No further modifications were made. The glass sandwich positioned at the front of the MaPMT, by means of a mechanical frame. Compared to the solution of gluing or using optical gels facilitates the disassembly for repair as well as improves the testing capabilities when choosing a different diameter of the air gap between scintillator and the MaPMT.

Table 1. Selected requirements for the S-DAM consisting of frontend and controler board.

| S-DAM Requirements | |
|---|---|
| **Component/Description** | **Requirement** |
| Connectivity Interface | Ethernet |
| Power Input | Single voltage connector to controller board |
| Dimensions | 50x50 mm$^2$, stackable |
| Time resolution | > 50 ns on each channel |
| Data output upstream | Each detected event with timestamp and channel number in standard mode. Energy value with timestamp and channel number for each channel in calibration mode. |
| Data output bandwidth | 50.000 detected events per second per module. S-DAM is capable to transfer all detected events. |

## Dedicated counting electronics

The dedicated electronics consist of a frontend module that houses the ASICs connected to the MaPMT and the control module. The control module houses the logic and performs pulse discrimination and processing. , It has the connections to the network and power supply. A selected list of the specific requirements can be found in Table 1.

It is important to note that this design can later transferred to a larger system for the upscaling process, where the control module not only supplies one frontend board but several without major modifications. This is one of the approaches to keep the power consumption low.

Another aspect that covers the commercial applicability of these modules is the possibility to supply the single modules with power over Ethernet. This way, encapsulated modules can be used independently for neutron or gamma investigations, in rough areas with a minimum of cables needed.

## Conclusion

A working 2x2 demonstrator was presented (see Fig. 1). This demonstrator will later used for further testing under working conditions in neutron scattering environments. These tests will help to prepare the upscaling of the SoNDe technology to a full size detector both in terms of interoperability of the modules as well as the know-how about handling the incoming data,adapted firm- and software.

The 2x2 demonstrator enables us to show the feasibility of all components of the SoNDe detector technology, such as modularity, interconnectivity and synchronicity between the single modules. Thus, it is a major step towards a full-scale detector following the same concept.

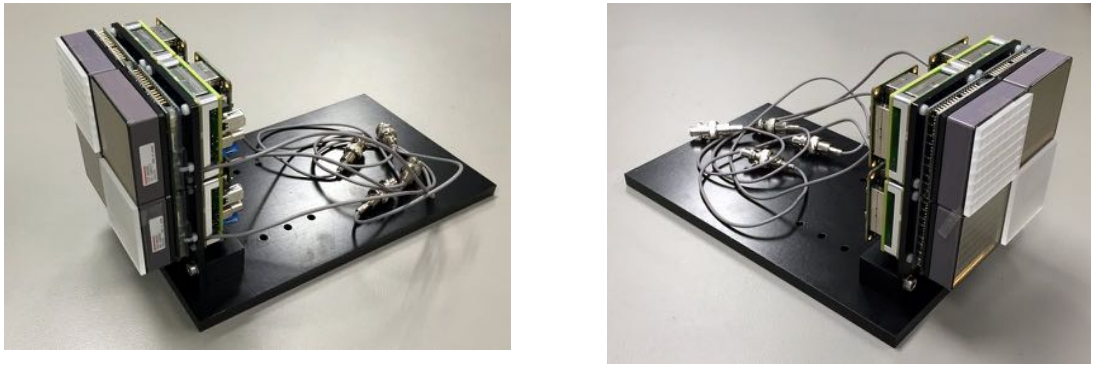

Fig. 2. Images of a mounted 2x2 demonstrator module. Different scintillator glasses are mounted for testing and comparison. The controller board has all connectors to the backend for data and electricity. The cables shown are for the high voltage supply of the MaPMTs.

This project is funded by the Horizon 2020 Framework Programme of the European Union. Project number 654124.

# SoNDe

## Solid-State Neutron Detector

### INFRADEV-1-2014/H2020

### Grant Agreement Number: 654124

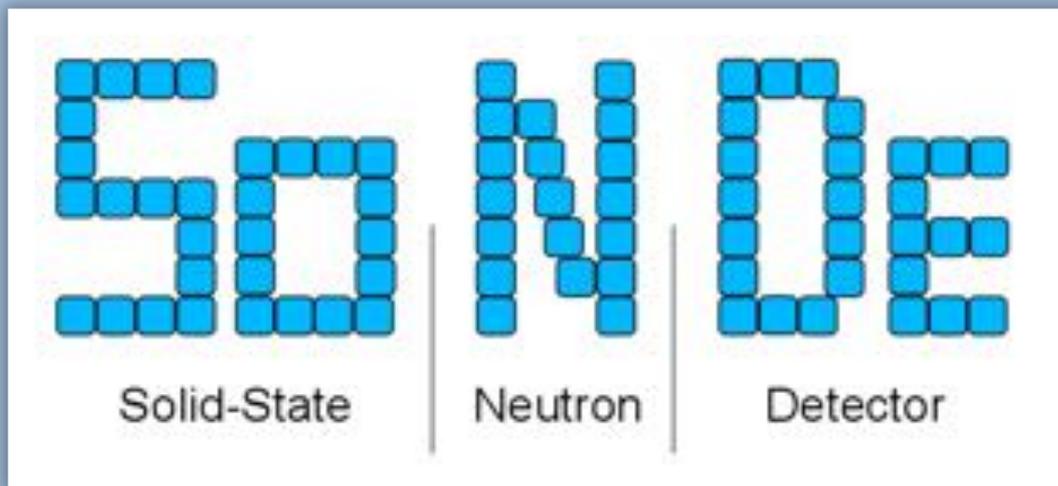

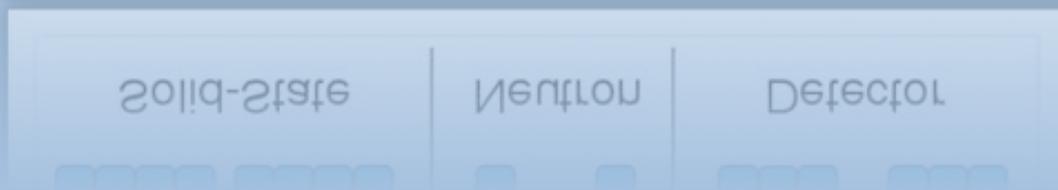

Deliverable Report:

D2.2 Test and Evaluation of 2x2 Demonstrator

This project is funded by the Horizon 2020 Framework Programme of the European Union. Project number 654124.

## Project and Deliverable Information Sheet

| SoNDe Project | |  |
|---|---|---|
| | Project Ref. No. 654124 | |
| | Project Title: Solid-State Neutron Detector SoNDe | |
| | Project Website: http://www.fz-juelich.de/ics/ics-1/DE/Leistungen/ESS/SoNDe-Projekt/ | |
| | Deliverable ID: D2.2 | |
| | Deliverable Nature: Report | |
| | Deliverable Level: PU | Contractual Date of Delivery: 31.01.2017 |
| | | Actual Date of Delivery: 16.05.2017 |
| | EC Project Officer: Darko Karacic | |

## Document Control Sheet

| Document | | |
|---|---|---|
| | Title: Solid-State Neutron Detector SoNDe | |
| | ID: 2x2-Demo-Test-Deliverable-D2.2 | |
| | Version: 1.0 | |
| | Available at: Project Website | |
| | Software Tool: MS Word 2011 | |
| | Files: 2x2-Demo-Test-Deliverable-D2.2.docx | |
| Authorship | Written by | Sebastian Jaksch, FZJ |
| | Contributors | Ralf Engels, Günter Kemmerling (FZJ), Codin Gheorge (IDEAS) |
| | Reviewed by | Sebastian Jaksch, FZJ |
| | Approved | Sebastian Jaksch, FZJ |

This project is funded by the Horizon 2020 Framework Programme of the European Union. Project number 654124.

## List of Abbreviations

| | |
|---|---|
| FZJ | Forschungszentrum Jülich, Jülich Research Centre |
| JCNS | Jülich Centre for Neutron Science |
| LLB | Laboratoire Léon-Brillouin |
| ESS | European Spallation Source |
| IDEAS | Integrated Detector Electronics AS |
| MaPMT | Multi-anode Photomultiplier Tube |
| S-DAM | SoNDe Data Acquisition Module |
| ROSMAP | IDEAS code name for counting electronics |

# Table of Contents

**This project is funded by the Horizon 2020 Framework Programme of the European Union. Project number 654124.**

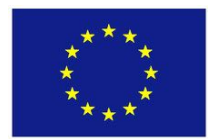

# Executive Summary

The previously presented 2x2 demonstrator [1] has to be tested thoroughly in order to make sure all requirements are met before an upscaling can be approached. These tests include both the performance of the demonstrators in terms of neutron countrate, in terms of electronic signal processing and in terms of usability.

These three fields were approached either with direct testing under neutron irradiation or on electronic test beds, where synthetic signals were used to test the behavior of the unit. The usability tests were limited to the development of a control software, which was necessary to perform the hardware tests.

In total all requirements were met or surpassed, which means both the development as well as the operation of the 2x2 demonstrators was an outstanding success. These results in turn allow us to proceed to the next project stage, where the technology that is working in a demonstrator is upscaled for use, both in larger numbers and for larger areas.

# Introduction

The SoNDe 2x2 Demonstrator is comprised of four identical submodules, each of which is performing all detection, analysis and communication functions on its own. For communication they are equipped with an Ethernet interface that allows bidirectional communication (control data, measurement data) as well as arranging several modules into an array using industrial Ethernet connections.

The functions that need to be performed in any single module for neutron detection are (1) scintillation conversion of a neutron event to a photon event, (2) detection of the photons within the MaPMT, (3) ADC conversion of the recorded photon energy, (4) discrimination of the ADC values in order to count only the desired events and no background (such as gamma radiation or high energy cosmic radiation), (5) allocating the detected event to a location (pixel) and time, (6) normalize all incoming data to a previously determined normalization procedure and (7) packing the data into and Ethernet/UDP package to transport the data to a host system where the data is either stored or can then be processed.

After that procedure the recorded data from the detector can be analyzed on a conventional computer without any need for data treatment. Thus these 2x2 Demonstrator modules allow a simple access to radiation detection with hand-held size technology.

In this document the evaluation of those modules is separated into electronic validation and validation under neutron irradiation.

# 1. EVALUATION OF ELECTRONIC PERFORMANCE

In this section mainly the electronic performance of the front-end component of the previously described 2x2 Demonstrator modules is evaluated. This means, from a function point of view that mainly the ADC conversion is performed in that module. For these evaluations mainly synthetic electronic test pulses from pulse generators were used in order to have a stable baseline for the evaluation of the electronic performance, as opposed to

This project is funded by the Horizon 2020 Framework Programme of the European Union. Project number 654124.

statical events. The effect of statistical distribution is taken into account by the evaluation under irradiation.

## 1.1. Specifications and Requirements

The specifications for the module are tailored to fit the output of a Hamamatsu H8500c MaPMT tube. This influences the dynamic range, the mechanical layout of the interface, the necessary gain compensation as well as the maximum current that can be expected during operation.

Additional free parameters are external requirements, such as the maximum acceptable power consumption or the timing accuracy for the detected events. Within limits also the maximum achievable counting rates are tunable parameters, albeit under the assumption that the MaPMT is not saturated.

The key parameters for the module are lists in Table 1.

**Table 1. Specification parameters for 2x2 SoNDe Demonstrator Module**

| Parameter | Value | Achieved |
|---|---|---|
| Interface | 64 Channel Hamamatsu H8500c interface | |
| Dynamic Range | 0 … - 80 pC | |
| Power | Max. 3 W per single module | |
| Event detection | Self-triggered for each channel | |
| Trigger threshold | -7 pC … 60 pC | |
| Dimensions | 52x52 mm$^2$ in order to fit seamlessly behind a H8500c | |
| Gain Compensation | Compensate for MaPMT gain variations (up to factor 3) | |
| Timing accuracy per event | <100 ns | |
| Communication | Ethernet | |
| Performance counting mode | Up to 100 kHz per channel | |
| Performance spectroscopic mode | 14 bit resolution for up to 50 kHz per channel | |
| Operating Conditions | Laboratory Conditions:<br><br>• Dry room<br>• Room temperature | |

This project is funded by the Horizon 2020 Framework Programme of the European Union. Project number 654124.

## 1.2. Dynamic Range Evaluation

In order to evaluate the dynamic range of the 2x2 demonstrator modules the single channels were tested with charge pulses between 9.4 and 94 pC, which is covering the required dynamic range well.

The signal is then fed through a preamplifier into an ADC delivering the digitized value for further analysis. The values measured after the preamplifier are given in Figure 1.

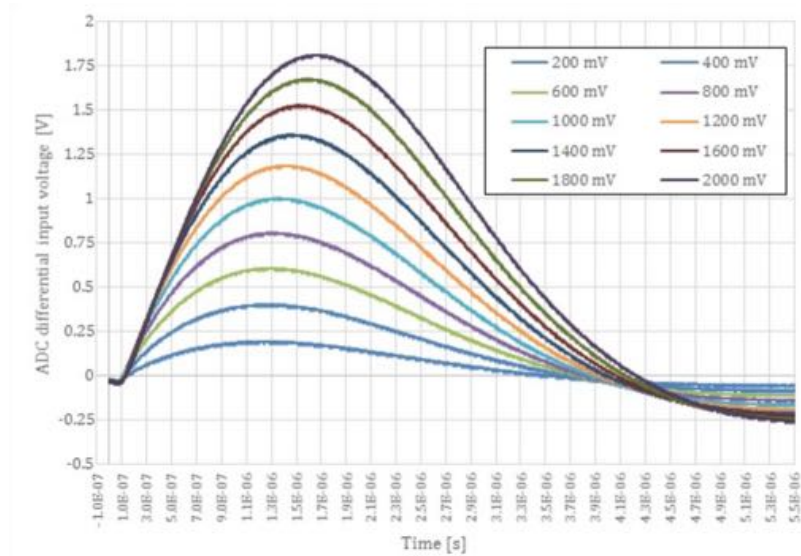

**Figure 1. Signal after preamplification before ADC conversion. The voltages correspond to event charges between 9.4 pC (200 mV) and 94 pC (2000 mV). The onset of the signal is stable in time and the shift in peak position and length of the pulse is continuous.**

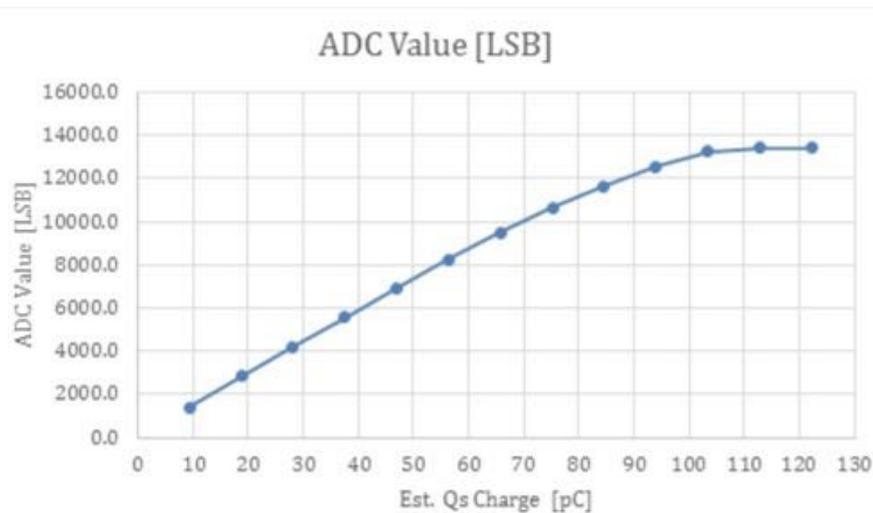

**Figure 2. ADC output value vs. estimated input charge. ADC value is well linear with the input charge up to 80 pC, meaning that a dynamic range up to 80 pC can be realized.**

**This project is funded by the Horizon 2020 Framework Programme of the European Union. Project number 654124.**

The resulting ADC values are given in Figure 2. There it is clearly visible that the behavior between the charge input and the ADC value is linear up to 80 pC. That linear behavior is necessary in order to perform an exact discrimination of the ADC values in the later logic components of the 2x2 demonstrator.

As the gain compensation is performed later in the logic components of the 2x2 demonstrator, the gain compensation using this ADC value is not limited by electrical constraints of the demonstrators and thus the dynamic range requirement is fulfilled.

### 1.3. Count Rate and Event Detection

The count rate of the modules for the single channels is limited by the time delay between input pulse and the length of the individual pulse in the detection system.

In order to evaluate the first contribution the single channels where triggered and the output of the trigger channels was monitored. The resulting trigger signals are shown in Figure 3. As the resulting trigger signal is consistently delayed by 110 ns after the event that triggered at, also for several channels. This means that the signals for different channels can be consistently timed without systematic errors between the single channels. However this is only true in the case of identical trigger charges.

The effect of different trigger values is illustrated in Figure 4.

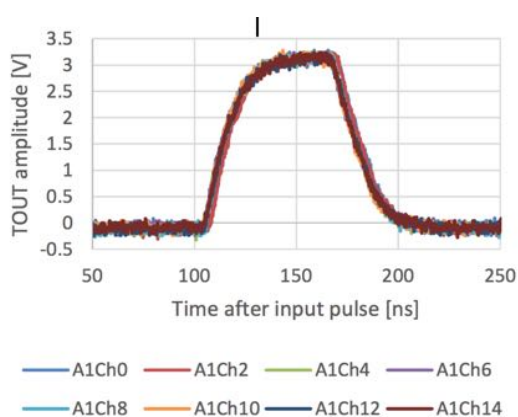

Figure 3.Time delays for triggering on different channels of a 2x2 demonstrator module. The time delay is constant at 110 ns, irrespective of the channel.

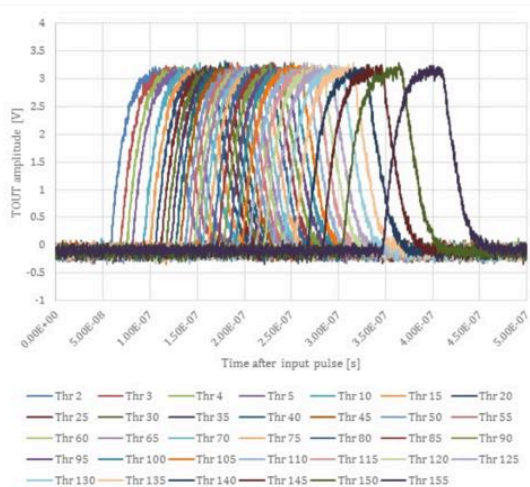

Figure 4. Time delay for different threshold settings. The lowest threshold settings are triggered faster, as it takes less time for the required amount of charge to be collected.

This project is funded by the Horizon 2020 Framework Programme of the European Union. Project number 654124.

It is immediately apparent that lower threshold values lead to a faster detection of the signal. This is an expected behavior as the higher charge of the threshold first has to be accumulated in order to reach the trigger threshold.

A similar behavior can be observed in the case of different excitation charges. Also here, lower charges take longer to reach the threshold, while higher charges are detected faster. The behavior is shown in.

These two effects have to be corrected for later data treatment.

The last effect of incoming pulses on the pulse rate is given by "lost pulses" in the case of a very high event frequency. The frequency for equidistant pulses, where pulses get lost was determined to be 2.23 MHz, while at 2.22 MHz a stable operation was still possible. This coincides well with the length of the trigger output (approximately 110 ns). Put back to back this pulses could create a frequency of approximately 9 MHz. As the charge in the single channels needs to decay before the next event can be triggered it is reasonable to assume that there has to be a dead time between the pulses, in this case 300 ns.

As the requirement in Table 1 was designed for statistically distributed signals the 2.22 MHz counting capability are misleading. However, using a dead time of 10 % the 2x2 demonstrator modules can still achieve approximately 200 kHz. So, even if the later data treatment in the module is imposing an additional delay the count rate requirement can be considered achieved.

## 2. Evaluation under Irradiation

In this section the overall 2x2 demonstrator (MaPMT, Front- and Backend) is evaluated. The test was performed at the TREFF instrument at FRM2 at a neutron wavelength of 4.7 Å. An image of the setup is shown in Figure 5.

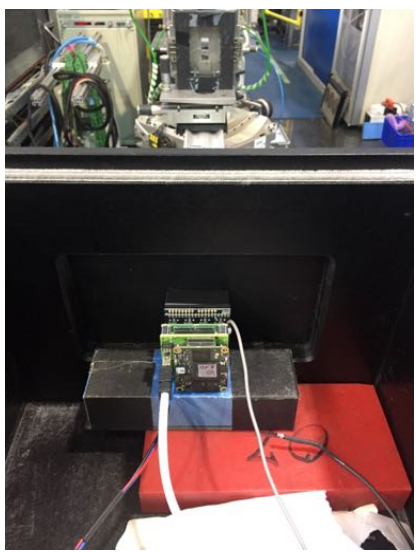

**Figure 5. Setup for testing a submodule of the 2x2 demonstrator under irradiation. The back of the module is visible, including power supply, Ethernet cable and HV cable. The neutrons are coming from the front, where on the top of the image the neutron guide is visible.**

This project is funded by the Horizon 2020 Framework Programme of the European Union. Project number 654124.

During those tests both the performance under real-life conditions as well as the usability of the software was tested. One of the interesting questions here was, whether the performance of the front-end board, that allows to meet the requirements also translates into a high performance of the over-all system. Additionally this offered the possibility to evaluate the control software, which will be needed for the ongoing testing.

The testing beam was constricted by two blinds to form a vertical stripe on the detector. This was done to make sure that in the inactive areas no event is triggered, which would lead to background in later application of the module. An image of the detected neutron events on the demonstrator surface is shown in Figure 6.

There it is apparent, that indeed the inactive areas remain dark. Additionally, there is no effect of saturation visible. The recorded pulse height spectrum allows a clear discrimination between neutron and gamma events.

All that information can be visualized and recorded in the control software (screenshot Figure 6). There the location of the events as well as the pulse height spectra is visible. Additionally each submodule of the 2x2 demonstrator can be accessed in a separate register card (here only one submodule is connected).

These test show that the performance of the 2x2 demonstrator under real live conditions (in the primary beam of a neutron scattering instrument) is indeed fulfilling all expectations. Under the circumstances the development of SoNDe can now proceed to the upscaling of the technology to larger areas.

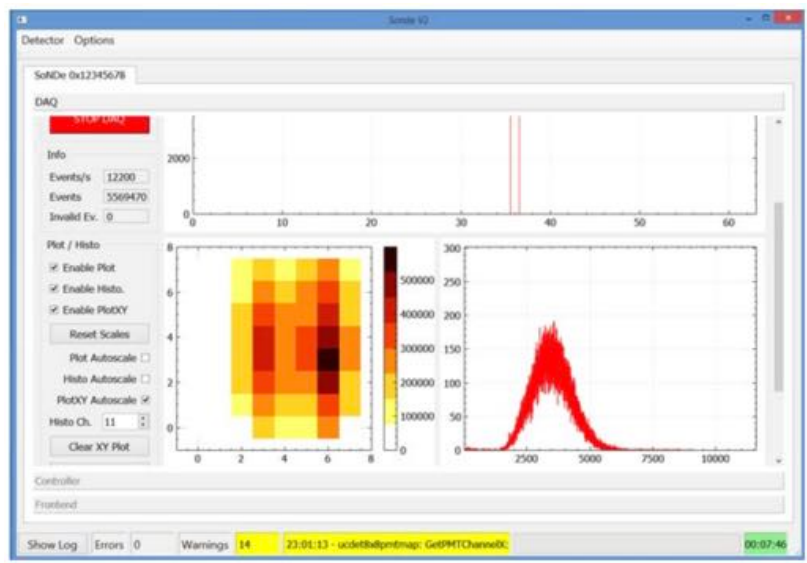

**Figure 6. Screenshot of the control and measurement software. The image of the detected neutrons (left image), shows the vertical slit of the beam on the detector surface. The pulse height spectrum (right image) only shows one single peak, which means that the discrimination of gammas is working and only neutrons are detected. On the top, the register for the respective module can be selected. If several modules are connected, each can be addressed by a single register in the software.**

This project is funded by the Horizon 2020 Framework Programme of the European Union. Project number 654124.

## 3. CONCLUSION

The evaluations presented above, both electronically and under application fulfill the requirements that were formulated at the beginning of the project.

That in turn allows to proceed to the upscaling of the technology. As the overall technology could be kept extremely compact during the development, also other fields of application open up, that were discussed in the additional applications report [2].

With the achievement of this deliverable also the milestone concept validation is achieved.

**This project is funded by the Horizon 2020 Framework Programme of the European Union. Project number 654124.**

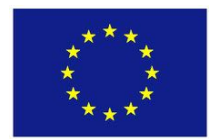